\documentclass[twocolumn,prb, letterpaper,showpacs,preprintnumbers,amsmath,amssymb]{revtex4-1}

\usepackage{graphicx}
\usepackage{dcolumn}
\usepackage{bm}
\usepackage{epsfig}
\usepackage{hyperref}

\begin{document}

\title{Combined cluster and atomic displacement expansion for solid solutions and magnetism}

\author{Kevin F. Garrity}
\email{kevin.garrity@nist.gov}
\affiliation{%
Material Measurement Laboratory, National Institute of Standards and Technology, Gaithersburg MD, 20899
}%

\date{\today}

\begin{abstract}
Finite temperature disordered solid solutions and magnetic materials
are difficult to study directly using first principles calculations,
due to the large unit cells and many independent samples that are required. In
this work, we develop a combined cluster expansion and atomic
displacement expansion, which we fit to first principles energies,
forces, and stresses. We then use the expansion to calculate
thermodynamic quantities at nearly first principles levels of
accuracy. We demonstrate that by treating all the relevant degrees of
freedom explicitly, we can achieve improved convergence properties as
compared to a simple cluster expansion, and our model naturally
includes both configurational and vibrational entropy. In addition, we
can treat coupling between structural and chemical or magnetic degrees
of freedom. As examples, we use our expansion to calculate properties
of Si$_{1-x}$Ge$_x$, magnetic MnO, Al with vacancies, and
Ba$_x$Sr$_{1-x}$TiO$_3$.
\end{abstract}

\maketitle

\section{\label{intro}Introduction}

Solid solutions, which are materials that have well-defined crystal
structures but disordered occupancy of atomic positions,
are important for a variety of technological applications as both structural and functional materials\cite{matsci_book}. About
half of the Inorganic Crystal Structure Database consists of compounds
with partial occupancy\cite{icsd}. Similarly, compounds with spin degrees of
freedom are often disordered at experimentally relevant temperatures. Unfortunately,
both of these types of materials are difficult to treat at a first
principles level of accuracy. Large supercells and averages over many
configurations are needed to treat disorder systematically, but the
computational cost of plane-wave density functional theory (DFT)
calculations increases rapidly with the number of atoms in a
calculation\cite{sqrs}. Even worse, finite temperature properties require
averages over thousands of steps of atomic motion.

Cluster expansions, which consist of models where chemical or spin
degrees of freedom are treated as interacting scalar variables on a
lattice, with all atomic displacements relaxed, are widely used to map
out the finite temperature phase diagrams of alloys and solid
solutions, as well as spin systems\cite{cluster_expansion,
  alloy_review, atat0, atat1, cluster_review}. However, simple cluster
expansions have several deficiencies. First, the atomic displacement
relaxations needed to fit cluster expansions are often computationally
expensive, as they require calculating the energies, forces, and
stresses of many intermediate structures during a relaxation, but only
the final energy is used in the model. Second, because atomic
displacements and strains are treated implicitly, models often
require effectively long-range and high-order interactions between
cluster variables, even if the underlying physical interactions are
short-ranged\cite{cluster_strain1, cluster_strain2,
  sige_relax_cluster, sige_relax_cluster2}.  Third, because they
eliminate all atomic degrees of freedom, simple cluster expansions do
not capture the effects of vibrational free energy, and attempts to
add vibrational free energy can be computationally
expensive\cite{alloy_review, cluster_vibrational,
  vibrational_ni3al}. Finally, because they eliminate all structural
information, simple cluster expansions can only calculate a very
limited number of properties. In particular, interactions between
structural and chemical degrees of freedom cannot be treated easily,
which excludes technologically relevant materials like
piezoelectrics, ferroelectrics, ferroelastics, magnetocalorics,
etc. that involve coupled degrees of freedom.

In this work, we combine the framework of a cluster expansion for
chemical or spin degrees of freedom with an atomic displacement
expansion, which has long been used to calculate finite temperature
properties of crystalline materials. Atomic displacement expansions up
to harmonic order, \textit{i.e.} phonon calculations, are routinely done either
as finite differences calculations or using DFT perturbation
theory\cite{dft-pt}, and higher order calculations are used to treat
anharmonic properties like thermal conduction or phase
transitions\cite{keivan_anharm, forceconst_alloys, compressed_sensing,
  lattice_dynamics, mechanically_unstable_model,
  second_principles, shengbte}. In contrast to some similar works on model
Hamiltonians\cite{batio3_modelham, batio3_modelham2,modelham_quantum}
that are sometimes used to treat solid solutions\cite{pzt_modelham,
  bst_modelham}, in this work, we keep all atomic displacement degrees
of freedom rather than only those related to soft modes, allowing us
to calculate more general properties.

By combining a cluster expansion with an atomic displacement
expansion, including interactions between them, we get a model with
many desirable properties. First, unlike models with specific
physically-inspired energy terms, our expansion can be easily applied
to any crystal structure or chemistry. Second, with only slight
variations, the same framework can treat chemical disorder, vacancies,
and magnetic disorder. Third, the model can be fit using standard
linear least-squares fitting techniques. Fourth, the model is
systematically improvable. Fifth, the model can be coupled to external
fields. Finally, because the model
naturally uses all of the energies, forces, and stresses from any
reference calculation and makes the relevant degrees of freedom
explicit, it can be fit with a relatively small number of DFT
calculations.

We have made the code to fit our expansion to first principles
calculations and evaluate new structures available online at
\url{https://github.com/usnistgov/spring_cluster}.  The rest of the work is
organized as follows. In Sec.~\ref{model}, we describe the expansion
form, symmetry properties, and fitting procedure we use in this
work. In Sec.~\ref{sec:examples}, we fit the model to several example
systems: Si$_{1-x}$Ge$_x$, magnetic MnO, Al with vacancies, and
Ba$_x$Sr$_{1-x}$TiO$_3$. Finally, in Sec.~\ref{conclu}, we present our
conclusions.

\section{\label{model} Expansion Form and Fitting}

\subsection{Expansion}

Our model consists of a Taylor expansion around a high symmetry
reference structure in terms of both scalar degrees of freedom and
vector atomic displacements, including interactions terms between
them. In this section, we will treat the case of a solid solution with
a single type of atomic substitution, which is represented by a scalar degree of
freedom. Magnetic and vacancy cases will be examined in the following
sections. In our formulas, the subscript indices $i$, $j$, ... run
over the atomic sites of the high symmetry supercell that we expand
around, and the superscript indices $x$, $y$, ... run over Cartesian
directions.

Our model consists of three main terms: a cluster expansion, an atomic
displacement expansion, and interaction terms between the two.

\begin{eqnarray}\label{eq:cluster}
E_{tot} = E_{cluster} + E_{atom} + E_{inter} 
\end{eqnarray}

The form of the cluster expansion is well-known:
\begin{eqnarray}\label{eq:cluster}
  E_{cluster} &=& \sum_i J_i s_i + \frac{1}{2!} \sum_{ij} J_{ij} s_i s_j \nonumber\\
  & &+ \frac{1}{3!} \sum_{ijk} J_{ijk} s_i s_j s_k + ... ,
\end{eqnarray}
where $s_i=0,1$ are scalar degrees of freedom at site $i$, with 1
corresponding to a dopant atom being present, and $J_{ij}$, etc., 
represent fitting coefficients. In contrast to a normal cluster
expansion, these energy terms represent the energy of
dopant atoms in the unrelaxed high symmetry reference structure, not
relaxed structures. Instead, we treat atomic displacements explicitly as follows:
\begin{eqnarray}\label{eq:cluster}
  E_{atom} &=& \frac{1}{2!} \sum_{ij}^{xy} K_{ij}^{xy} u_i^x u_j^y \nonumber\\
  & &+ \frac{1}{3!} \sum_{ijk}^{xyz} K_{ijk}^{xyz} u_i^x u_j^y u_k^b + ...,
\end{eqnarray}
where $u_i^x$ is the displacement of atom $i$ in direction $x$ from
its reference position. $K_{ij}^{xy}$ is the fitting coefficient for
the interaction between atom $i$ moving in the $x$ direction and atom
$j$ moving in the $y$ direction; other terms are similar. There is no
first-order term because we assume the high symmetry structure is in
equilibrium. The second order term is the standard
harmonic force constant matrix, and higher order terms are anharmonic
force constants. Forces are obtained by taking a derivative with
respect to $u_i^x$ in the normal fashion: $F_i^x = -\partial E_{tot}
/ \partial u_i^x$.

Finally, we include interaction terms between the scalar and vector degrees of freedom:
\begin{eqnarray}\label{eq:cluster}
  E_{inter} &=& \sum_{ij}^{x} M_{ij}^{\,x} s_i u_j^x 
  + \frac{1}{2!}\frac{1}{1!} \sum_{ijk}^{x} M_{ijk}^{\,\ x} s_i s_j u_k^x \nonumber\\& & +  \frac{1}{1!}\frac{1}{2!} \sum_{ijk}^{xy} M_{ijk}^{\,xy} s_i u_j^x u_k^y + ...,
\end{eqnarray}
where $M_{ij}^{\,x}$, etc. are fitting coefficients for the
interaction terms. For example, the first-order term in this expansion, with coefficient $M_{ij}^{\,x}$,
is turned on if there is a dopant at site $i$ ($s_i=1$), and
determines the forces on the surrounding atoms $j$ in direction $x$ that result from that
substitution. Similarly, the term $M_{ijk}^{\,xy}$ represents the
change the spring constant between the atoms at sites $j$, $k$ in
directions $x$, $y$ due to a dopant at site $i$.

This expansion is very general and can in principle be used for any
combination of substitutions and atomic distortions that maintains the
topology of the bonding in the crystal structure. While the expansion
must be truncated in practice, it can be systematically improved if
higher precision is needed. We will demonstrate in
Sec.~\ref{sec:examples} that is also useful in practice, and as
discussed in Sec.~\ref{sec:sige} this expansion will often have better
convergence properties than an expansion that treats some of the
degrees of freedom implicitly.

\subsection{Symmetry}

While the above expansion can in principle handle any reasonably small
distortion of a unit cell, the number of fitting coefficients
increases rapidly as higher-order terms are needed. To make the scheme
useful, it is necessary to take advantage of symmetries of the
reference structure in order to reduce the number of independent
fitting coefficients. We will present a brief overview of the symmetry
properties; most properties carry over from discussions of atomic displacement expansions.\cite{keivan_anharm, anharmonic}

The energy must be invariant under the application of the space group
symmetries of the high symmetry reference structure, which consist of a 
symmetry matrix $R^{xy}$ and potentially a partial translation $\tau^x$. Under
the application of a symmetry operation, the site $i$ can be shifted
to another site $i'$: $X_{i'}^x = \sum_y R^{xy} X_i^y +
\tau^x$, where $X_i^x$ is the reference position of atom $i$.  Because
the invariance must hold for any combination of $s_i$ and $u_i^x$,
each term in our expansion must be individually invariant. For
example:

\begin{eqnarray}\label{eq:sym}
J_{ij} &=& J_{i'j'} \\
K_{ij}^{xy} &=& \sum_{zw} R^{xz} R^{yw} K_{i'j'}^{zw} \\
M_{ijk}^{\,\ x} &=& \sum_{y} R^{xy} M_{i'j'k'}^{\ \ \ y}.
\end{eqnarray}

These relations are widely known and used for separate cluster and
atomic displacement expansions, and aside from keeping track of which
degrees of freedom transform as scalars and which as vectors, the
relations in this work are analogous.

In addition to space group operations, the energy must be invariant
under permutations of either the cluster degrees of freedom or the
displacement degrees of freedom\cite{keivan_anharm, anharmonic}. For example:
\begin{eqnarray}
  J_{ij} &=& J_{ji} \\
  K_{ij}^{xy} &=&  K_{ji}^{yx} \\
M_{ijk}^{\,\ x} &=& M_{jik}^{\,\ x}\label{eq:permute},
\end{eqnarray}
We note that scalar and vector degrees of
freedom cannot be permuted for each other.

In addition to space group operations, each term in the model must
also be invariant under arbitrary translations of the unit cell\cite{keivan_anharm, anharmonic}. These
relations are also known as acoustic sum rules because of their role in ensuring that there are three zero
frequency phonon modes at $\Gamma$. The acoustic sum rules for our
expansion are again simple generalizations of the relations from pure atomic
displacement expansions. For example:
\begin{eqnarray}\label{eq:asr}
0&=&\sum_{j} K_{ij}^{xy} \quad \forall \, i,xy \label{eq:asr1}\\
0&=&\sum_{j} M_{ij}^{\,x} \quad \forall \, i,x \\
0&=&\sum_{k} M_{ijk}^{\,xy} \quad \forall \, ij,xy \label{eq:asr2}
\end{eqnarray}

There are similar constraints due to the
invariance of the system under arbitrary rotations of the unit cell\cite{anharmonic,keivan_anharm}.
These additional constraints relate different orders of the expansion
to each other; however, we do not enforce them explicitly during our
fitting procedure.

\subsection{Strain}

In addition to cluster and atomic displacement variables, it is
necessary to include strain degrees of freedom, $\epsilon^{xy}$, in
our model.  However, strains are fundamentally related to long
wavelength atomic displacements, and our existing expansion does not
require any new fitting coefficients to treat
strain\cite{anharmonic,kun_huang}. The relationship between the
harmonic force constants and the elastic constants $C^{wx,yz}$ is
well-known, albeit rarely used in first principles contexts:
\begin{eqnarray}
E_{strain} &=& \frac{1}{2} \sum^{wx,yz} C^{wx,yz} \epsilon^{wx} \epsilon^{yz} \label{strain1} \\
S^{wx,yz} &=& \frac{1}{2} \sum_{ij} K_{ij}^{wx} (X_i^{y} - X_j^{y})(X_j^{z} - X_i^{z}) \\
C^{wx,yz} &=& S^{wy,xz} + S^{xy,wz} - S^{xw,zy}
\end{eqnarray}
where $X_i^x$ is the reference position of atom $i$ in direction $x$,
and $S^{wx,yz}$ is a tensor defined above. Elastic constants have an
extra permutation relation, $C^{wx,yz}=C^{yz,wx}$, that in some cases
results in an additional constraint on the force constants.  We
enforce this relation by requiring that the spring constants obey the
Kun-Huang condition, $S^{wx,yz} = S^{yz,wx}$\cite{anharmonic,
  kun_huang}.

The contributions to the elastic constants due to
dopants are treated using analogous formulas, except with $s_i$
variables that turn on the extra contributions in the presence of
dopants. For example:
\begin{eqnarray}
E_{cl-strain} &=& \frac{1}{2} \sum_i^{wx,yz} C_i^{\,wx,yz} s_i \epsilon^{wx} \epsilon^{yz} \\
S_i^{\,wx,yz} &=& \frac{1}{2} \sum_{jk} M_{ijk}^{\,wx} (X_j^{y} - X_k^{y})(X_j^{z} - X_k^{z}) \\
C_i^{\,wx,yz} &=& S_i^{\,wy,xz} + S_i^{\,xy,wz} - S_i^{\,xw,zy}
\end{eqnarray}

In addition to the above terms, which are second order in strain, there are additional first order in strain effective
interactions. The lowest order interaction term between pure atomic displacements and strain, $E_{at-strain}$, is:
\begin{eqnarray}
E_{at-strain} &=& \sum_{i}^{xyz} T_{i}^{x,yz} u_i^x \epsilon^{yz} + ...  \\
T_{i}^{x,yz} &=& \sum_{j} K_{ij}^{xy} (X_j^{z} - X_i^{z}),\label{T}
\end{eqnarray}
where $T_{i}^{x,yz}$ is the first order coupling between strain
$\epsilon^{yz}$ and the atomic displacement $u_i^x$. We emphasize that
$T_{i}^{x,yz}$ is fully determined by appropriate combinations of our
existing coupling coefficients and is not an independent fitting
parameter. Similarly, the lowest order coupling
between a cluster variable $s_i$ and strain $\epsilon^{xy}$, is a
simple generalization of Eq.~\ref{T}:
\begin{eqnarray}
  E_{cl-strain} &=& \sum_{i}^{xy} U_{i}^{\ xy} s_i \epsilon^{xy} \nonumber \\
 & &+ \sum_{i}^{xy} V_{ij}^{\ x,yz} s_i u_j^x \epsilon^{yz} + ...\\
  U_{i}^{\ xy} &=& \sum_{j} M_{ij}^{\,x} (X_j^{y} - X_i^{y}) \\
  V_{ij}^{\ x,yz} &=& \sum_{j} M_{ijk}^{\,xy} (X_k^{z} - X_j^{z}),\label{strainend}
\end{eqnarray}

By including all terms of Eqs.~\ref{strain1}-\ref{strainend}, we
include the effects of strain up to second order; we ignore higher
order strain contributions. Stress is calculated in the normal way,
$\sigma_{xy} = -\frac{1}{V} \partial E_{tot} / \partial
\epsilon_{ij}$.

\subsection{Vacancies}

We can represent vacancies (or interstitial atoms) using the same
formalism as discussed above, with non-zero cluster degrees of freedom
representing missing atoms instead of substituted atoms. The only
difficulty is that unless we impose additional constraints, the energy
will depend weakly on the displacement of a vacancy site, which is
unphysical as there is no atom to displace. The necessary constraints
simply require the force on any vacancy site to be exactly zero. These
constraints enforce a cancellation between various terms in the model
to ensure the energy does not depend on vacancy positions. These
constraints can be constructed naturally by the same procedures used
to setup the linear regression.

\subsection{\label{mag} Magnetism}

We can use our expansion to treat simple magnetic systems, with the
spins represented by cluster variables. If the spins are limited to
collinear up and and down spins, and have nearly constant magnitude,
then our expansion still applies. The only difference is that the
cluster variables become $s_i=\pm 1$, as in the Ising model. A
collinear magnetic field can be treated as a chemical potential.

Unlike cluster variables, which represent different atoms, there is an
additional symmetry in spin systems due to the invariance of the
energy when flipping every spin ($s_i \rightarrow -s_i$). This symmetry
requires that expansion terms have an even number of spin variables. 

In many cases, an Ising-like model will be inadequate to treat the
magnetic degrees of freedom. A natural expansion of this formalism is
to use the more general Heisenberg model, where the spins are allowed
to rotate in three dimensions and magnetic anisotropy can be
included. A full model would require treating the spin degrees of
freedom like vectors with constant magnitude, instead of
scalars. However, for simple situations, it is possible to fit the
expansion to simple collinear spins, but then allow the spins to
rotate during calculations using the model.  This can be done by
treating interactions between pairs of scalar spins like a dot product
($s_i s_j \rightarrow \vec{s}_i \cdot \vec{s}_j$). Using this idea, we
can fit our expansion to collinear spin data using the Ising-like
expansion, but allow for Heisenberg-like spin-spin interactions when
calculating finite temperature properties.

\subsection{\label{fit}Fitting}

Despite its generality, our model is linear in the coupling
coefficients, and can be fit using standard linear least-squares
techniques. We first use Gaussian elimination to determine the
minimum number of independent fitting parameters after applying
Eqs. \ref{eq:sym}-\ref{eq:permute}.  We prepare a library of
energies, forces, and stresses from a set of DFT calculations that are
relevant to the desired application. Then, we fit the coefficients,
enforcing the acoustic sum rules (Eqs. \ref{eq:asr1}-\ref{eq:asr2}) as linear constraints.

In order to decide how to truncate our expansion, we use recursive
feature elimination\cite{recursive_fe}, with cross-validation to determine how many
independent parameters to keep for optimum out-of-sample
prediction. During each step of this algorithm, the smallest
standardized coefficients are dropped and the model is refit. We find
that this procedure is faster and more numerically stable than $L_1$
regularization, which has previously been used to search for sparse models in a similar context\cite{compressed_sensing}.

We use cutoff distances to limit the initial set of fitting parameters
to a reasonable number, with high order terms given shorter
cutoffs. Typically, the lowest order terms are cutoff only by the size of
the supercell, but we limit anharmonic terms to second or third
nearest neighbor interactions.

For insulating materials, electrostatic dipole-dipole interactions decay
very slowly with distance, and it is necessary to handle these contributions
seperately from our fitting procedure. We use Born effective charges and the electronic dielectric
constant, determined from DFT perturbation theory\cite{dft-pt, gonze},
to subtract the contribution of long-range electrostatic forces before
fitting. We then add back the long-range contribution for
predictions. For the current examples where atoms of the same valence
are substituted, we assume the Born effective charges are not modified
by substitutions, atomic displacements, or strain.

In order to test and improve our models, we also implement the ability
to use automatically run finite temperature Monte Carlo sampling to
generate new structures, perform new DFT calculations on these
structures, and add the new data to the model in order to iteratively
improve it. This technique is especially useful when the initial model
is unstable at finite temperature. In this case, the model will escape
the physically relevant phase space and develop a large negative
energy\cite{second_principles}. After an instability develops, we can
automatically identify a structure near the instability, run a new DFT
calculation, and include the new data point in our model.  Continuing
in this fashion, we can iteratively improve the model, continuing
until it is stable and reaches the desired accuracy.

\subsection{First principles computational details}

We perform first principles DFT calculations\cite{hk,ks} with a
plane-wave basis set as implemented in QUANTUM ESPRESSO\cite{QE} and
using the GBRV high-throughput ultrasoft pseudopotential
library\cite{ultrasoft,gbrv}. We use a plane wave cutoff of 40 Ryd.
We use the PBEsol exchange-correlation functional\cite{pbesol}, which
provides more accurate lattice constants than other generalized
gradient approximation functionals. For Mn $d$ states, we use DFT+U
with U=3 eV. We calculate Born effective charges and dielectric
constants using DFT perturbation theory\cite{dft-pt}. We take
advantage of the computational efficiency of using non-diagonal
supercells to calculate long-range two-body model terms\cite{nondiag}.

\subsection{Monte Carlo Sampling}

In order to determine finite temperature properties using our model,
we perform classical Monte Carlo sampling of the Boltzmann
distribution using the Metropolis algorithm\cite{montecarlo}.  We use
a local updating strategy, performing sweeps where we attempt to move
each atom a random distance, the average magnitude of which is tuned
during thermalization to achieve approximately 50\% acceptance.
Strain is treated in a similar fashion. In cases where we also allow
the cluster variables to change, we perform grand canonical Monte
Carlo, doing additional cluster sweeps using a single spin flip
approach.

\section{\label{sec:examples}Examples}

\subsection{\label{sec:sige}Si$_{1-x}$Ge$_x$}

As our first example, we consider the technologically relevant solid
solution Si$_{1-x}$Ge$_x$, in the diamond structure. The end members
are the only thermodynamically stable phases at zero temperature.

\begin{figure}
\includegraphics[width=3.3in]{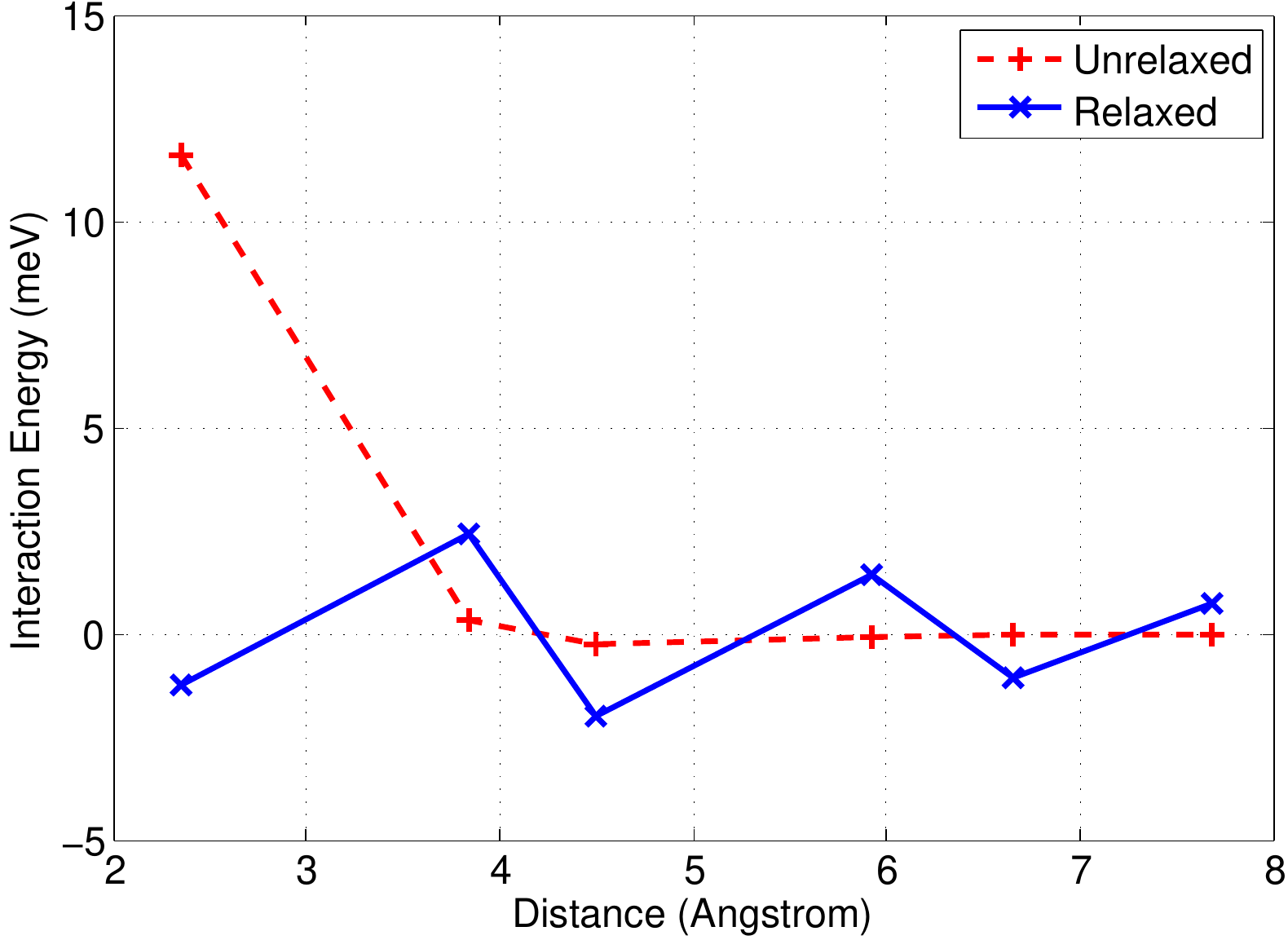}
\caption{\label{fig:sige0} Interaction energy between two Ge substituted for Si as a function of distance. Red dashed line is for unrelaxed positions, blue solid line is for relaxed positions.}
\end{figure}

We begin by investigating the effects of atomic relaxation on the
convergence of interactions in
Si$_{1-x}$Ge$_x$\cite{sige_relax_cluster}. We consider a
$4\times4\times4$ cell of Si (128 atoms) with two Ge atoms substituted
for two Si. In Fig.~\ref{fig:sige0} we plot the energy difference
between a given configuration of Ge atoms, as a function of distance
between them, taking as the reference an isolated Ge atom, either
relaxed or unrelaxed, in the same cell. In the case where all the
atoms are fixed to their ideal positions, the first neighbor
effective interaction between the Ge is large, but decays to zero rapidly as the
distance increases. Even the second neighbor direct interaction is
almost negligible. In contrast, when all the atoms in the cell are
allowed to relax, the first neighbor interaction energy is
smaller, but it barely decays with distance. A cluster expansion fit
to the relaxed energies will have much worse convergence with distance
than a fixed atom cluster expansion. By taking into account atomic
displacements explicitly, we can take advantage of this improved
convergence. We typically find that the limiting factor in our
expansions is the convergence of the force constants, rather than
the pure cluster terms. Also, we find that in the fixed atom
case, three-body cluster interactions are either
very short-ranged or negligible.

Moving on to fitting our expansion, we first fit to structures with
only small atomic displacements. Our fitting data consists of
structures with pure Si, pure Ge, and random substitutions of Si and
Ge, as well as random atomic displacements of up to 0.15 \AA. We use
40 calculations to fit, with supercell sizes of up to $4 \times 4
\times 4$ the primitive unit cell (128 atoms), and we test our
expansion with another 30 structures, some of which have $2 \times 2
\times 8 $ unit cells to test possible longer range interactions. Our
model allows interactions up to second order in the cluster variables
and up to third order in the atomic displacements, although the third
order terms are only nearest-neighbor. We note that because the highest
order terms in our model are an odd power of the atomic displacements, the model cannot be applied
for arbitrarily large displacements, but it can successfully reproduce
the properties of the material within its range of validity.

\begin{figure}
\includegraphics[width=3.3in]{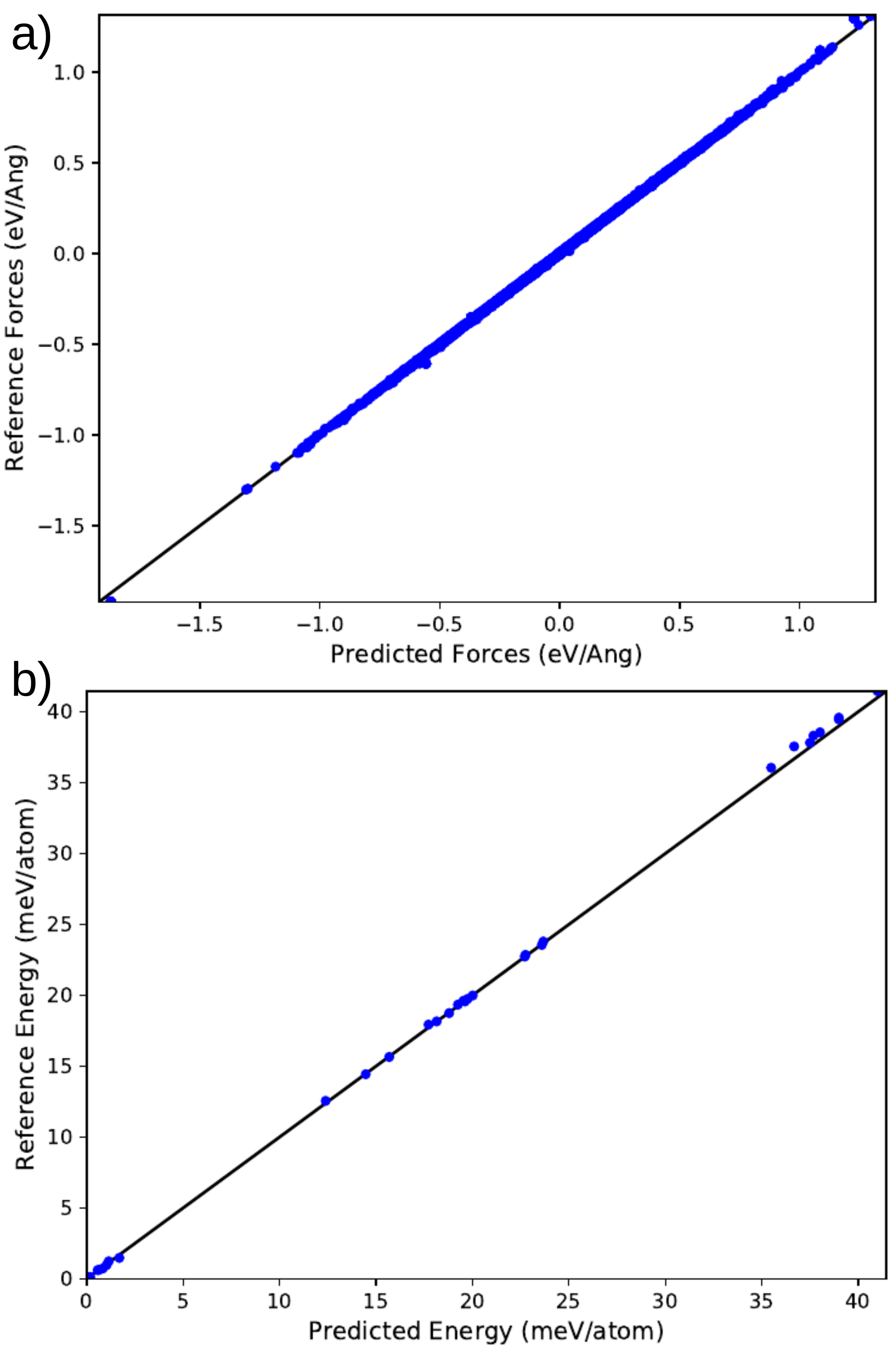}
\caption{\label{fig:sige1} For small distortions of the Si$_{1-x}$Ge$_x$ system, out-of-sample comparison between model (x-axis) and first principles (y-axis) a) forces (eV/\AA) and b) energies (meV/atom).}
\end{figure}

After performing recursive variable selection and cross-validation, we
are left with 95 independent fitting parameters. We show the
out-of-sample forces and energies predicted by the model in
Fig.~\ref{fig:sige1}. We get excellent agreement over a considerable
range of energies and forces, with a mean absolute error in energies
of 0.2 meV / atom, or 1.0\%, and a mean absolute error in force
components of 1.4\%.

\begin{figure}
\includegraphics[width=3.3in]{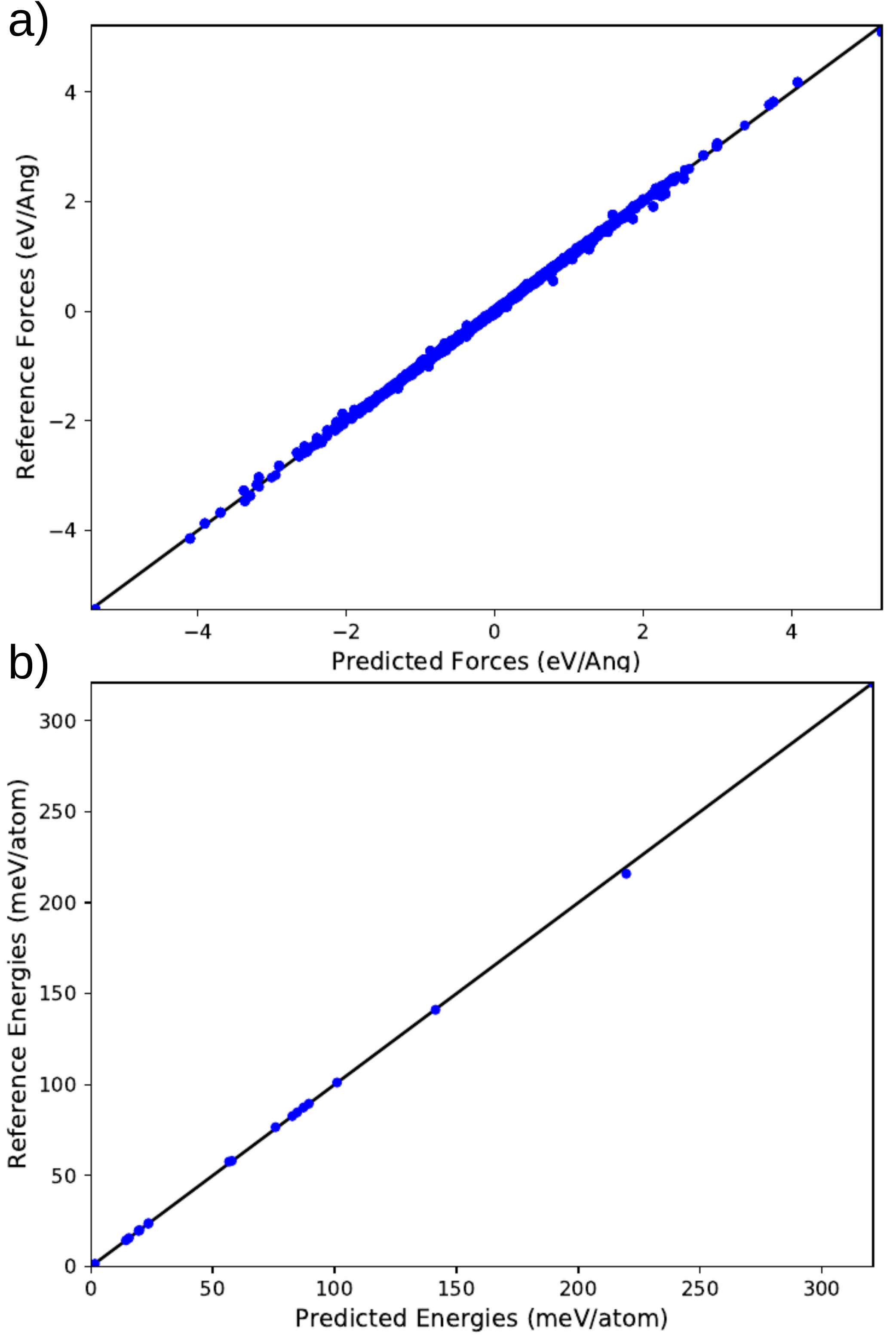}
\caption{\label{fig:sige2} For larger distortions of the Si$_{1-x}$Ge$_x$ system, out-of-sample comparison between model (x-axis) and first principles (y-axis) a) forces (eV/\AA) and b) energies (meV/atom).}
\end{figure}

Next, we generate structures with larger distortions, up to 0.7 \AA,
and we allow expansion terms up to fourth order in the atomic
displacements. As shown in Fig.~\ref{fig:sige2}, our new model again
has excellent performance, with an out-of-sample mean absolute error
of 0.3 meV / atom, as compared to an average energy of 63 meV / atom
in the testing data.  The model is suitable for thermodynamic
calculations up to several hundred Kelvin and includes all relevant
anharmonic contributions to the energy.

\begin{figure}
\includegraphics[width=3.3in]{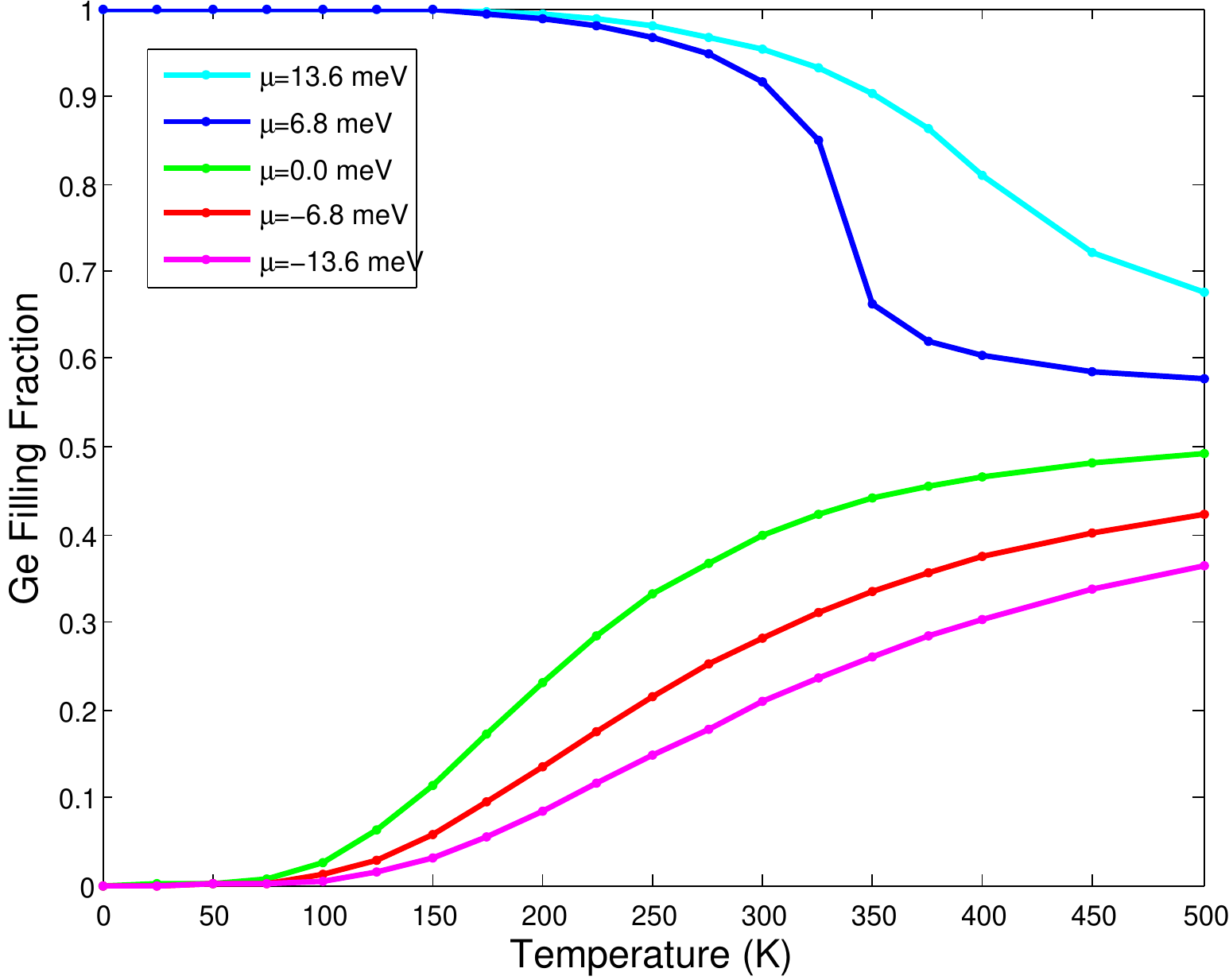}
\caption{\label{fig:sige_pd} Ge filling fraction as a function of
  temperature (K) for Si$_{1-x}$Ge$_x$, for fixed chemical
  potential $\mu$. Different color lines are different chemical potentials,
  see legend.}
\end{figure}

In Fig.~\ref{fig:sige_pd}, we show an example grand canonical Monte
Carlo calculation in a $10\times10\times10$ unit cell. For several
fixed chemical potentials, we plot the Ge filling fraction as a
function of temperature. At low temperature, the system is phase
separated into pure Ge and pure Si. Near 300 K, depending on the
chemical potential, the system transitions to a disordered solid
solution, in reasonable agreement with past
results on this system\cite{sige_relax_cluster, sige_old, atat0}.

\subsection{MnO}

MnO in the rocksalt structure has an antiferromagnetic (AFM) ground state
with a Neel temperature of 118 K\cite{mno}. Even without spin-orbit
coupling, the spin structure breaks cubic symmetry and leads to a
rhombohedral distortion of the unit cell of 0.96$^\circ$, according to
our calculations.  This distortion reduces the frustration of the
first nearest neighbor antiferromagnetic spin-spin interactions, only
half of which can be satisfied in the cubic structure\cite{mno}. The
second neighbor interactions are comparable in size to the first
neighbor interactions, and are not frustrated. We find that the energy
difference between the ferromagnetic (FM) and antiferromagnetic phases in
the cubic structure is 77 meV/Mn, and increases to 86 meV/Mn when the
rhombohedral distortion is relaxed; so, the distortion energy is
small, but not negligible.

\begin{figure}
\includegraphics[width=3.3in]{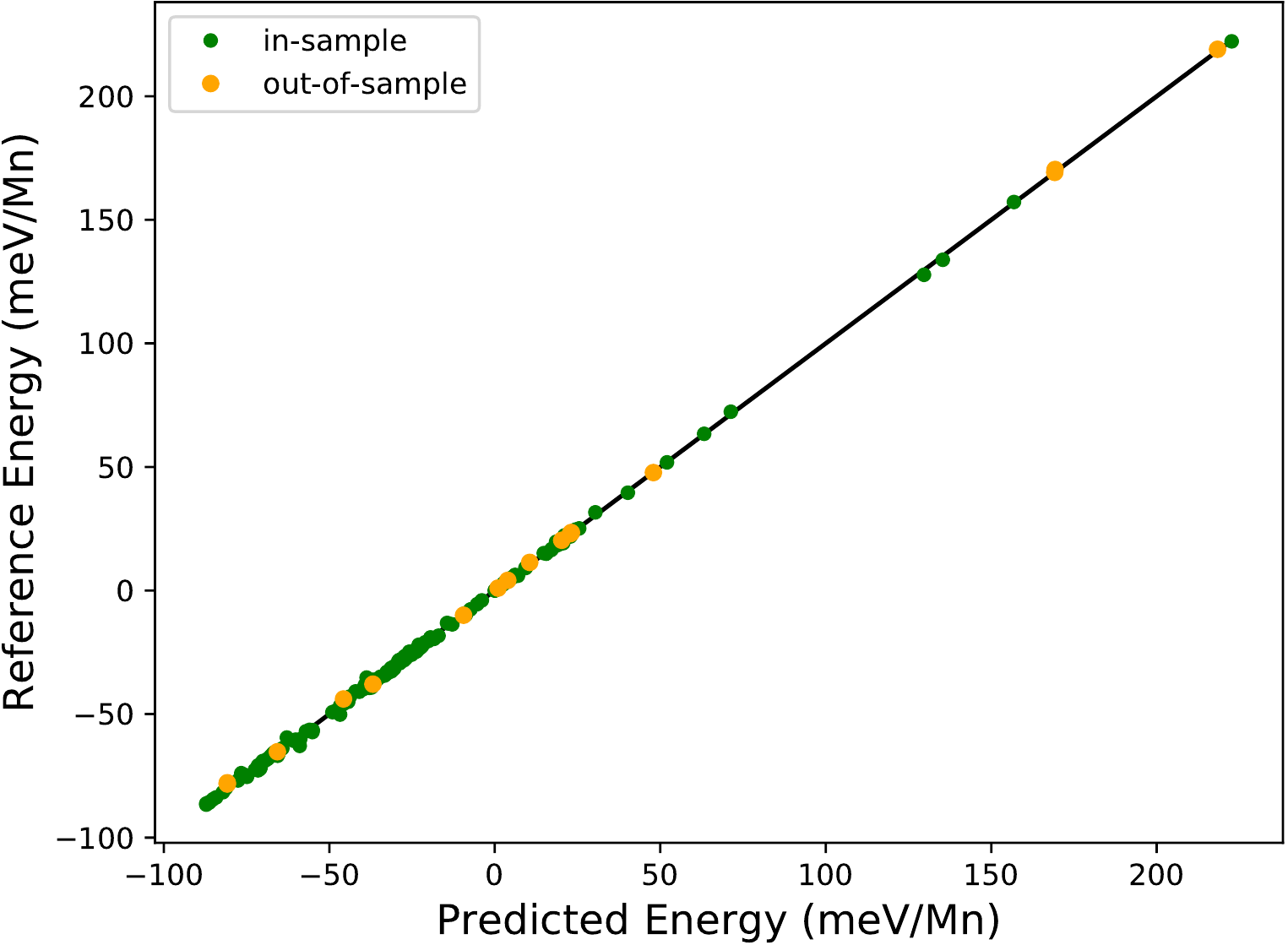}
\caption{\label{fig:mno_fit} For MnO, comparison between model
  energies and first principles energies, in meV/Mn. Green symbols are
  in-sample, larger orange symbols are out-of-sample. Reference energy is cubic MnO in FM phase.}
\end{figure}

We seek to study this coupled magnetic-structural phase transition
using our model. We expand around the cubic ferromagnetic
structure. We allow only second order terms in the spin degrees of
freedom, as discussed in Sec. \ref{mag}.  We consider first and second
order atomic displacement terms, which can be coupled to spin, as well
as short-range third and fourth order displacement terms. We fit the
model to a set of structures with both ordered and random collinear
spins, and atomic displacements up to 0.5 \AA in $4\times4\times4$ supercells.  In
Fig.~\ref{fig:mno_fit}, we show a comparison between our model and
reference DFT calculations. We find excellent agreement, with a mean
absolute energy error of 0.9 meV/Mn over a wide energy range.

\begin{figure}
\includegraphics[width=3.3in]{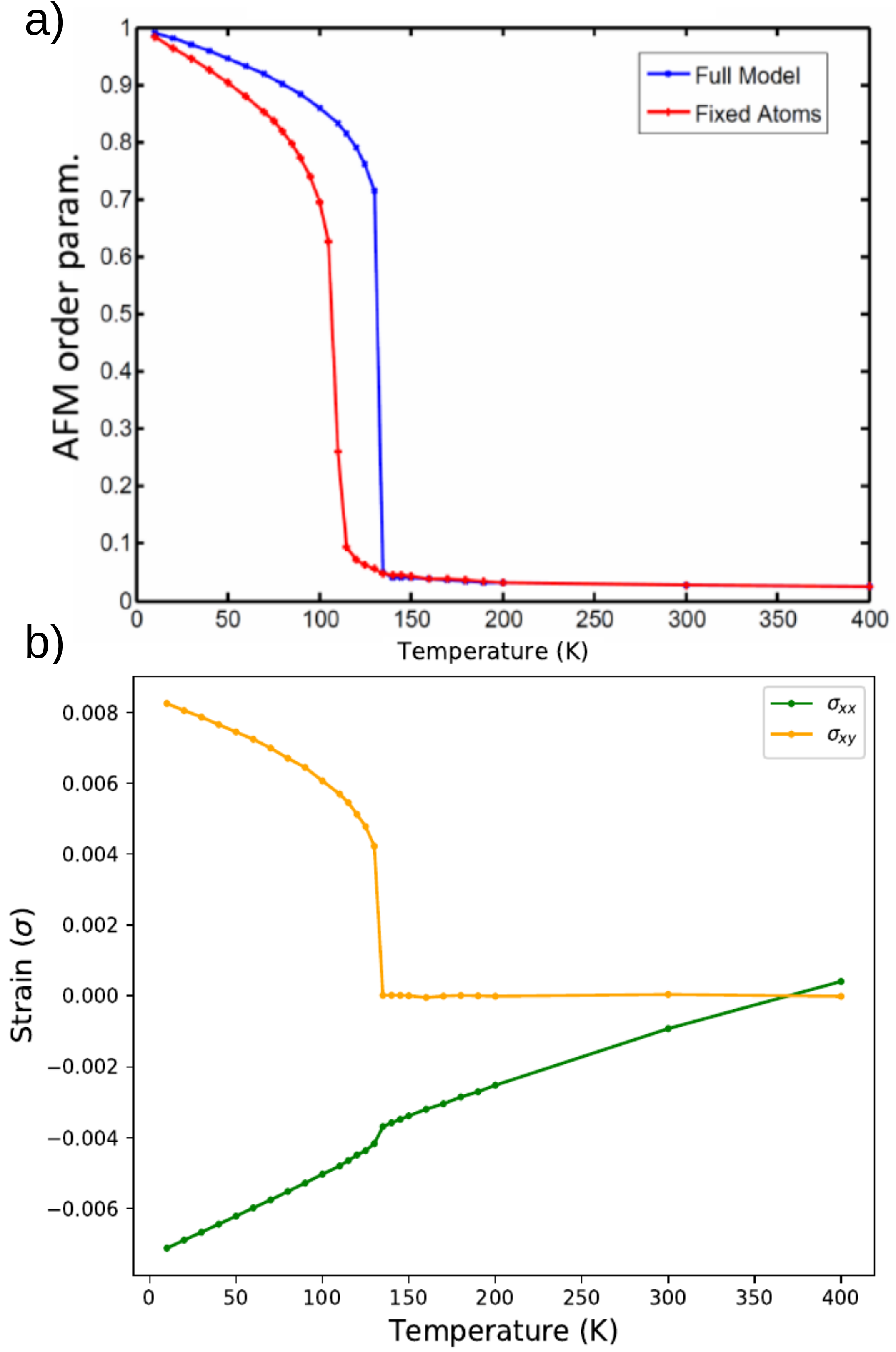}
\caption{\label{fig:mno_pd} a) In MnO, AFM order parameter as a
  function of temperature (K). Blue line allows full atomic
  displacements, red line fixes them. b) Strain vs. temperature
  (K). Green line is $\sigma_{xx}$, orange is $\sigma_{xy}$.}
\end{figure}

We use the model to preform classical Monte Carlo sampling in a $12
\times 12 \times 12$ unit cell, interpreting the spins as
Heisenberg-like, as discussed in Sec. \ref{mag}. In
Fig.~\ref{fig:mno_pd}a, we plot the antiferromagnetic order parameter
as a function of temperature, both using the full model and with the
atoms fixed to the cubic structure and only the spins allowed to
relax. The coupling between the spin variables and
structural variables both raises the phase transition temperature and
modifies its character, making the transition more strongly first
order. In addition, we can use the model to examine how the structure
changes near the phase transition. In Fig.~\ref{fig:mno_pd}a, we plot
both the $xx$ and $xy$ components of strain as a function of
temperature. As expected, the $xy$ component is only nonzero in the
low temperature AFM phase, and its magnitude is closely related to the
AFM order parameter. The $xx$ component is less affected by the
transition, but there is a minor change in slope as well as a small
volume jump at the phase transition. However, this volume change is much
smaller than the 2.2\% volume difference between the AFM and FM phases
at zero temperature, which emphasizes the fact that the cubic
paramagnetic phase at finite temperature is not well approximated by
the cubic FM phase at zero temperature.

We note that the coupled structural-magnetic finite temperature
calculations performed in this section are straightforward using our
expansion method, but would be challenging to calculate directly using
first principles techniques or using a pure spin-spin magnetic model.

\subsection{Al with vacancies}

\begin{figure}
\includegraphics[width=3.3in]{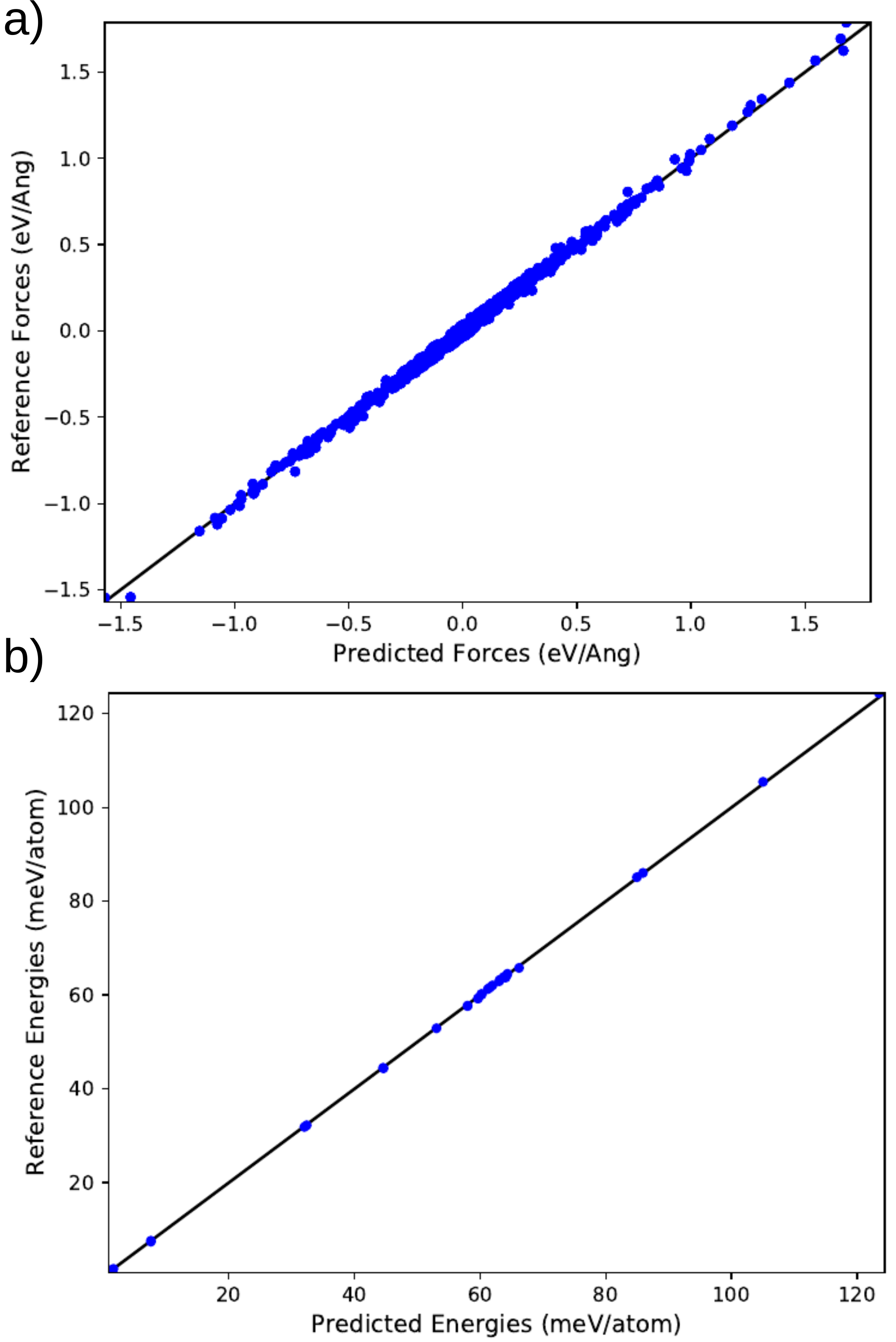}
\caption{\label{fig:al} For Al with vacancy system, out-of-sample comparison between model (x-axis) and first principles (y-axis) a) forces (eV/\AA) and b) energies (meV/atom).}
\end{figure}

As an example of a vacancy calculation, we consider Al in the
\textit{fcc} structure\cite{al}. We consider $3\times3\times3$ and
$6\times6\times6$ unit cells with 0-2 vacancies per cell (up to 7\%),
and atomic displacements of up to 0.26 \AA. The reference
chemical potential is that of zero temperature bulk Al. Our expansion
consists of terms up to fourth order in the atomic displacements and
up two second order in the cluster variables, although the third and
forth order terms are very short-ranged.

As shown in Fig.~\ref{fig:al}, our expansion gives excellent
out-of-sample agreement with the reference forces and energies. One
current limitation of the model is that it does not allow vacancies to
hop from site to site, which will begin to happen at fairly low
temperatures in Al. This limits the thermodynamic calculations we can
perform on this system to low temperatures. An extension of this
calculational framework to handle barrier hopping events is a possible
future direction of research.

\subsection{Ba$_x$Sr$_{1-x}$TiO$_3$}

BaTiO$_3$ and SrTiO$_3$ are well-studied perovskite oxides that are
used technologically for their dielectric
properties\cite{lattice_dynamics,second_principles,modelham_quantum,batio3_modelham2,batio3_modelham,
  bst_modelham}. At low temperatures, BaTiO$_3$ is a rhombohedral
ferroelectric, due to a polar distortion along the (111) direction. As
the temperature is raised, BaTiO$_3$ becomes orthorhombic, with
polarization along the (011) direction, then tetragonal, with
polarization along the (001) direction, and finally cubic.

SrTiO$_3$ is also cubic at high temperatures, but goes through at a
non-polar tetragonal phase transition related to octahedral rotations
($a^0a^0c^-$ in Glazer notation\cite{glazer}). As has been
well-studied, at zero temperature in DFT calculations, SrTiO$_3$ still
has a weak polar distortion, even after taking octahedral rotations
into account\cite{sto_rotations}. Zero temperature
quantum fluctuations, which we do not include in our model, are
necessary to get the correct ground state of SrTiO$_3$, and are
understood to be responsible for the enormous low temperature
dielectric constant of SrTiO$_3$\cite{modelham_quantum}.

\begin{figure}
\includegraphics[width=3.3in]{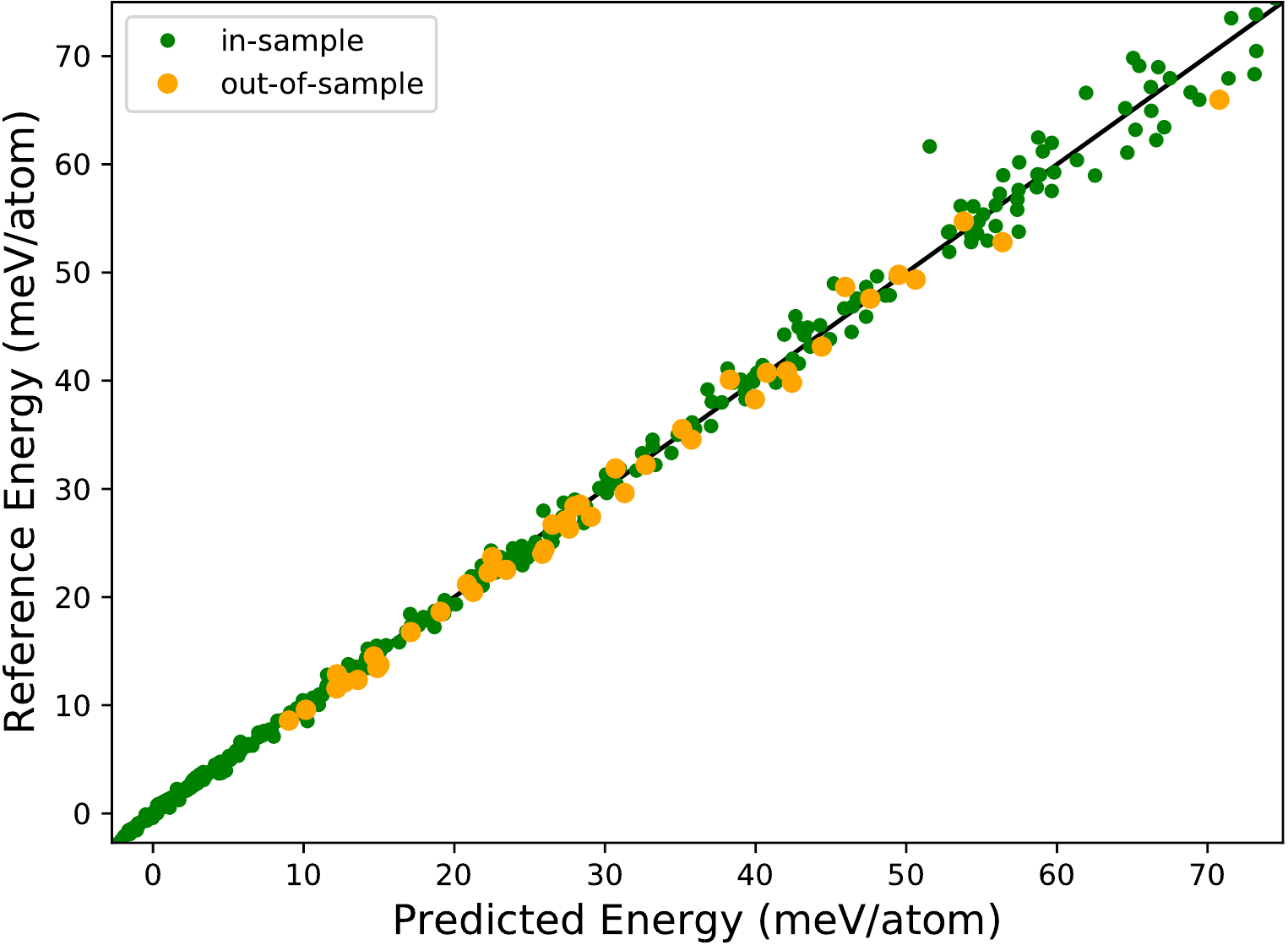}
\caption{\label{fig:bsto} For Ba$_x$Sr$_{1-x}$TiO$_3$, comparison between model
  energies and first principles energies, in meV/atom. Green symbols are
  in-sample, larger orange symbols are out-of-sample. }
\end{figure}

We fit our expansion to the Ba$_x$Sr$_{1-x}$TiO$_3$ system, expanding
around cubic BaTiO$_3$. Due to the fact that both end members are
unstable in their high-symmetry phases, this system requires a much
more careful treatment of the anharmonic modes than previous
examples. It is necessary to include DFT calculations from the various
locally stable minima in addition to the experimentally observed
structures in order to ensure that model gives good results at finite
temperature.  We find that we can get accurate results up to 300 K by
expanding only up to fourth-order in atomic displacements,
second-neighbor in distance, and including up to three-body
interactions. We use the recursive approach discussed in
Sec.~\ref{fit} to identify instabilities in the model and generate new
fitting data until the model reaches sufficient accuracy. In addition,
we weigh low energy structures more in our fitting to ensure the local
minima are well described.

As can be seen in Fig.~\ref{fig:bsto}, model performance is not quite
as good as the previous examples that lack unstable modes, and it
begins to degrade around 50 meV/atom. However, the mean absolute error
in energy is still only 1.4 meV/atom, as compared to an average energy
of 36.8 meV/atom in our test set. Furthermore, the model is more
accurate at lower energies, allowing it to describe local minima
correctly, which is necessary to capture low temperature behavior.

\begin{figure}
\includegraphics[width=3.3in]{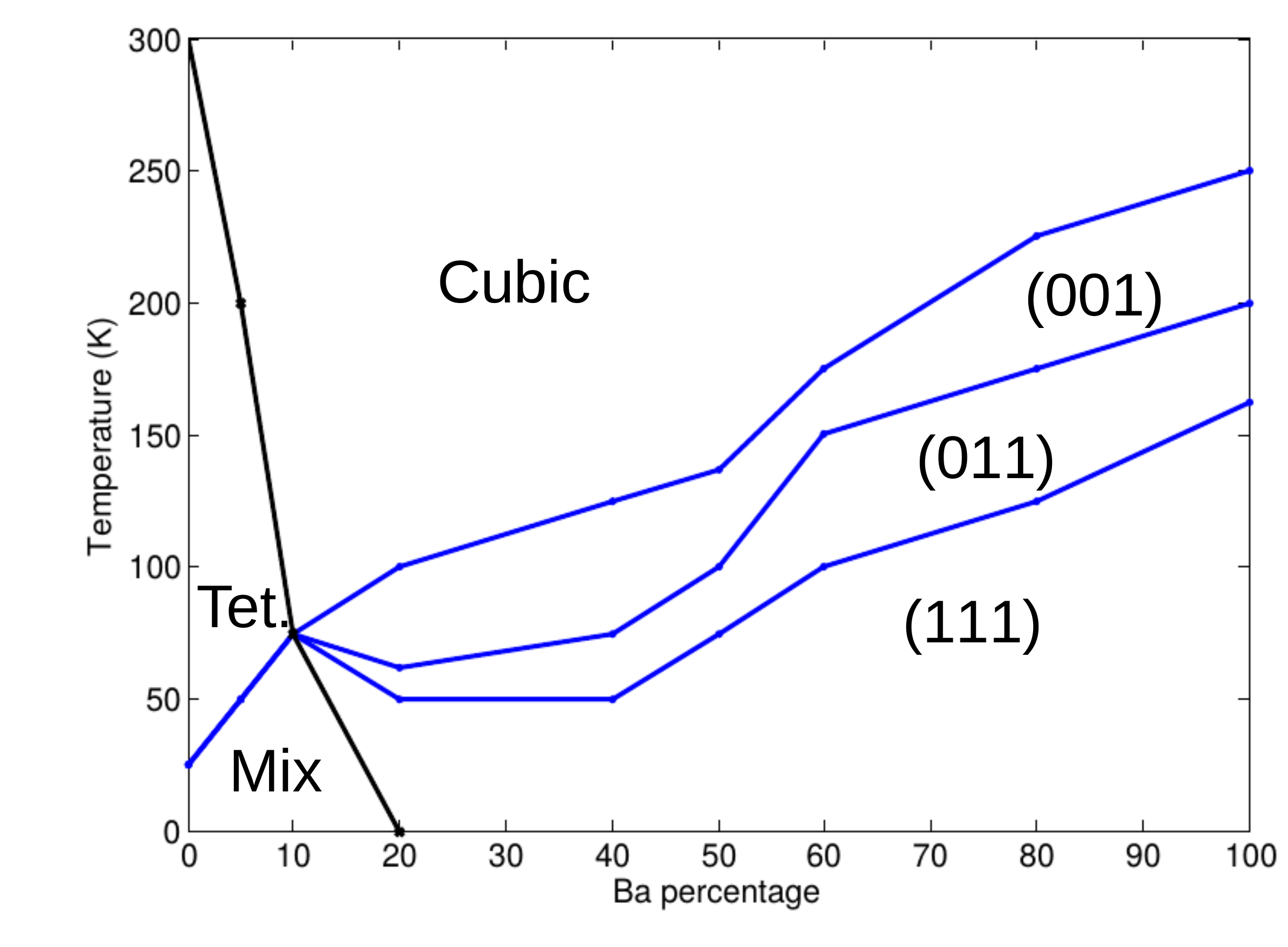}
\caption{\label{fig:pd} Phase diagram of
  Ba$_x$Sr$_{1-x}$TiO$_3$ as a function of doping and temperature (K). Tet refers to non-polar tetragonal, Mix
  refers to mixed octahedral rotations and polarization, and (111),
  (011) and (001) refer to polarization directions.}
\end{figure}

We use our model to run Monte Carlo calculations at finite
temperatures, with Sr and Ba distributed randomly, but fixed during
each calculation. We use a 10$\times$10$\times$10 unit cell. In
Fig. ~\ref{fig:pd}, we show the resulting phase diagram. We reproduce
the three ferroelectric phase transitions on the Ba-rich portion of
the phase diagram, as well as the non-polar phase transition on the
Sr-rich side. In agreement with previous first principles based
calculations in this
system,\cite{lattice_dynamics,second_principles,modelham_quantum,batio3_modelham2,batio3_modelham,
  bst_modelham} we overestimate the SrTiO$_3$ phase transition
temperature and underestimate the BaTiO$_3$ phase transition
temperatures. Considering the small energy differences involved and
the sensitivity of unstable modes to changes in volume, this level of
accuracy is typical for first principles phase diagrams. Due to the
fact that we do not include quantum fluctuations, we find that the low
temperature phase of SrTiO$_3$ is polar, instead of a quantum
paraelectric.

\begin{figure}
\includegraphics[width=3.3in]{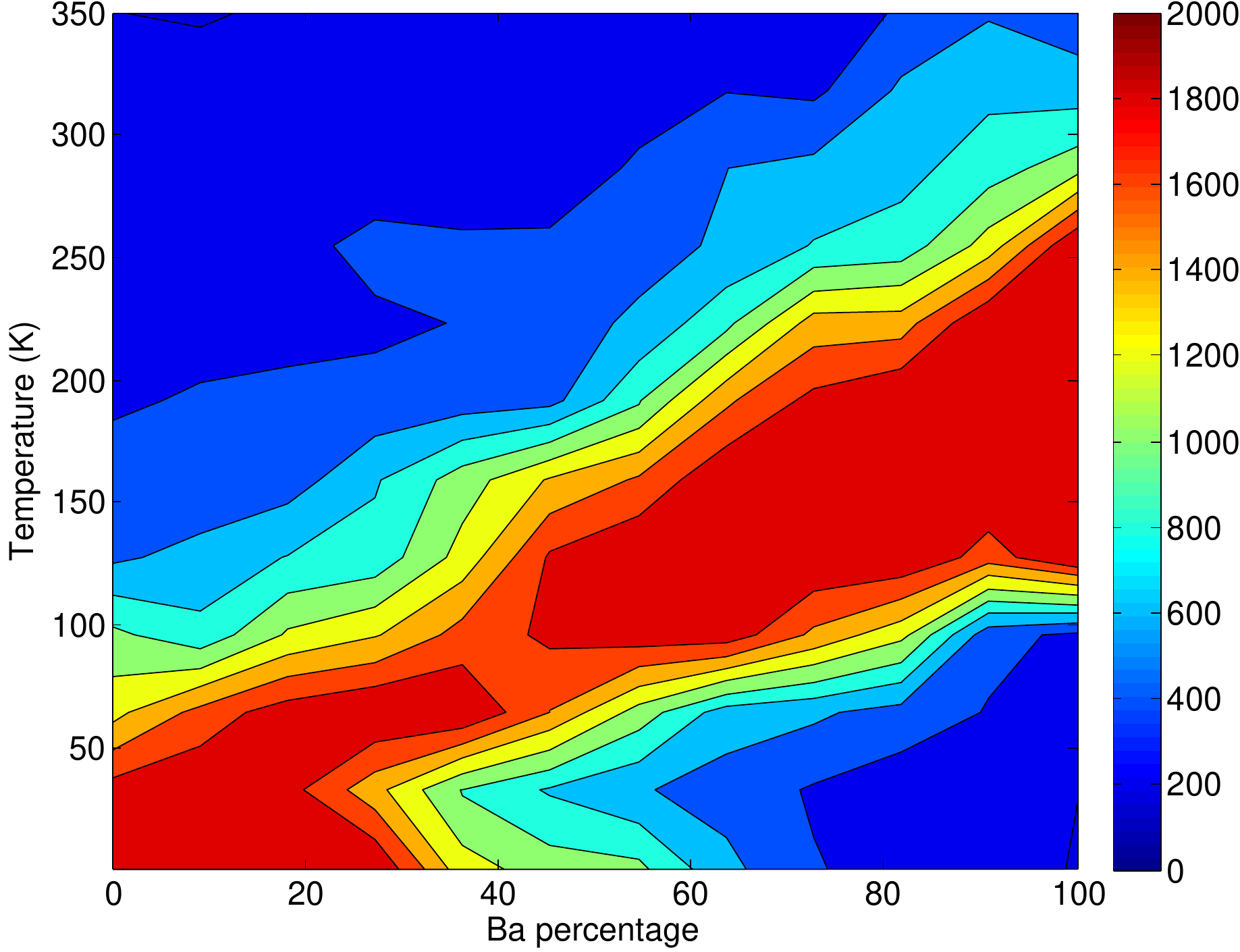}
\caption{\label{fig:diel} Static dielectric constant as a function of Ba percentage and temperature (K). Colors range from 0 (dark blue) to 2000 (dark red).}
\end{figure}

In addition to identifying the phases, we can use our model to
calculate detailed material properties as a function of doping and
temperature. For example, in Fig. ~\ref{fig:diel}, we present the
average dielectric constant throughout the phase diagram, calculated with the method of Ref. \onlinecite{pzt_modelham}. Unlike a
effective Hamiltonian approach\cite{pzt_modelham}, our expansion includes
contributions from all atomic degrees of freedom, instead of just soft
modes, and treats thermal expansion correctly. Due limitations in converging the dielectric constant near a
ferroelectric phase transition in a finite cell, we cap the reported
dielectric constant at 2000. As expected, the dielectric constant is
very high throughout the region where the various polar phase
transitions occur, but is unaffected by the non-polar transition in
the Sr-rich region.

\section{\label{conclu}Conclusions}

In this work, we present an expansion in terms of both scalar degrees
of freedom, corresponding to chemical or magnetic variables, and
vector atomic displacements. We discuss various symmetry properties of
the expansion, as well as a procedure for determining relevant
coefficients and fitting them to first principles calculations. By
using examples, we show that the model can be usefully applied to
solid solutions of semiconductors like Si$_{1-x}$Ge$_x$ and oxides
like Ba$_x$Sr$_{1-x}$TiO$_3$, as well as magnetic insulators like MnO
and metals with vacancies like Al.

Due to the fact that this expansion can be applied to a wide range of
materials and can be fit in a nearly automatic fashion, we expect that
it can be useful for many purposes. By combining features of
a cluster expansion with structural degrees of freedom, we can achieve
improved convergence with distance, and by making use of energies, forces, and
stresses from every available self-consistent field calculation, we
can fit the expansion with surprisingly few first principles
calculations. The expansion naturally includes both configurational
free energy and vibrational free energy to any desired order. In
addition, we can calculate properties that couple structural
properties with chemical or magnetic degrees of freedom. This allows
for the study of materials like ferroelectrics, piezoelectrics,
electrocalorics, magnetocalorics, shape memory alloys, and
ferroelastics that are technologically relevant but difficult to treat
directly with first principles calculations.

\begin{acknowledgments}
We wish to acknowledge discussions with Eric Cockayne and Kamal Choudhary.
\end{acknowledgments}


\begin{thebibliography}{44}%
\makeatletter
\providecommand \@ifxundefined [1]{%
 \@ifx{#1\undefined}
}%
\providecommand \@ifnum [1]{%
 \ifnum #1\expandafter \@firstoftwo
 \else \expandafter \@secondoftwo
 \fi
}%
\providecommand \@ifx [1]{%
 \ifx #1\expandafter \@firstoftwo
 \else \expandafter \@secondoftwo
 \fi
}%
\providecommand \natexlab [1]{#1}%
\providecommand \enquote  [1]{``#1''}%
\providecommand \bibnamefont  [1]{#1}%
\providecommand \bibfnamefont [1]{#1}%
\providecommand \citenamefont [1]{#1}%
\providecommand \href@noop [0]{\@secondoftwo}%
\providecommand \href [0]{\begingroup \@sanitize@url \@href}%
\providecommand \@href[1]{\@@startlink{#1}\@@href}%
\providecommand \@@href[1]{\endgroup#1\@@endlink}%
\providecommand \@sanitize@url [0]{\catcode `\\12\catcode `\$12\catcode
  `\&12\catcode `\#12\catcode `\^12\catcode `\_12\catcode `\%12\relax}%
\providecommand \@@startlink[1]{}%
\providecommand \@@endlink[0]{}%
\providecommand \url  [0]{\begingroup\@sanitize@url \@url }%
\providecommand \@url [1]{\endgroup\@href {#1}{\urlprefix }}%
\providecommand \urlprefix  [0]{URL }%
\providecommand \Eprint [0]{\href }%
\providecommand \doibase [0]{http://dx.doi.org/}%
\providecommand \selectlanguage [0]{\@gobble}%
\providecommand \bibinfo  [0]{\@secondoftwo}%
\providecommand \bibfield  [0]{\@secondoftwo}%
\providecommand \translation [1]{[#1]}%
\providecommand \BibitemOpen [0]{}%
\providecommand \bibitemStop [0]{}%
\providecommand \bibitemNoStop [0]{.\EOS\space}%
\providecommand \EOS [0]{\spacefactor3000\relax}%
\providecommand \BibitemShut  [1]{\csname bibitem#1\endcsname}%
\let\auto@bib@innerbib\@empty
\bibitem [{\citenamefont {Callister}\ and\ \citenamefont
  {Rethwisch}(2013)}]{matsci_book}%
  \BibitemOpen
  \bibfield  {author} {\bibinfo {author} {\bibfnamefont {W.~D.}\ \bibnamefont
  {Callister}}\ and\ \bibinfo {author} {\bibfnamefont {D.~G.}\ \bibnamefont
  {Rethwisch}},\ }\href@noop {} {\emph {\bibinfo {title} {Materials Science and
  Engineering: An Introduction, 9th Edition}}}\ (\bibinfo  {publisher}
  {Wiley},\ \bibinfo {year} {2013})\BibitemShut {NoStop}%
\bibitem [{\citenamefont {Belsky}\ \emph {et~al.}(2002)\citenamefont {Belsky},
  \citenamefont {Hellenbrandt}, \citenamefont {Karen},\ and\ \citenamefont
  {Luksch}}]{icsd}%
  \BibitemOpen
  \bibfield  {author} {\bibinfo {author} {\bibfnamefont {A.}~\bibnamefont
  {Belsky}}, \bibinfo {author} {\bibfnamefont {M.}~\bibnamefont
  {Hellenbrandt}}, \bibinfo {author} {\bibfnamefont {V.~L.}\ \bibnamefont
  {Karen}}, \ and\ \bibinfo {author} {\bibfnamefont {P.}~\bibnamefont
  {Luksch}},\ }\href {\doibase 10.1107/S0108768102006948} {\bibfield  {journal}
  {\bibinfo  {journal} {Acta Crystallographica Section B}\ }\textbf {\bibinfo
  {volume} {58}},\ \bibinfo {pages} {364} (\bibinfo {year} {2002})}\BibitemShut
  {NoStop}%
\bibitem [{\citenamefont {Zunger}\ \emph {et~al.}(1990)\citenamefont {Zunger},
  \citenamefont {Wei}, \citenamefont {Ferreira},\ and\ \citenamefont
  {Bernard}}]{sqrs}%
  \BibitemOpen
  \bibfield  {author} {\bibinfo {author} {\bibfnamefont {A.}~\bibnamefont
  {Zunger}}, \bibinfo {author} {\bibfnamefont {S.-H.}\ \bibnamefont {Wei}},
  \bibinfo {author} {\bibfnamefont {L.~G.}\ \bibnamefont {Ferreira}}, \ and\
  \bibinfo {author} {\bibfnamefont {J.~E.}\ \bibnamefont {Bernard}},\ }\href
  {\doibase 10.1103/PhysRevLett.65.353} {\bibfield  {journal} {\bibinfo
  {journal} {Phys. Rev. Lett.}\ }\textbf {\bibinfo {volume} {65}},\ \bibinfo
  {pages} {353} (\bibinfo {year} {1990})}\BibitemShut {NoStop}%
\bibitem [{\citenamefont {Sanchez}\ \emph {et~al.}(1984)\citenamefont
  {Sanchez}, \citenamefont {Ducastelle},\ and\ \citenamefont
  {Gratias}}]{cluster_expansion}%
  \BibitemOpen
  \bibfield  {author} {\bibinfo {author} {\bibfnamefont {J.}~\bibnamefont
  {Sanchez}}, \bibinfo {author} {\bibfnamefont {F.}~\bibnamefont {Ducastelle}},
  \ and\ \bibinfo {author} {\bibfnamefont {D.}~\bibnamefont {Gratias}},\ }\href
  {\doibase https://doi.org/10.1016/0378-4371(84)90096-7} {\bibfield  {journal}
  {\bibinfo  {journal} {Physica A: Statistical Mechanics and its Applications}\
  }\textbf {\bibinfo {volume} {128}},\ \bibinfo {pages} {334 } (\bibinfo {year}
  {1984})}\BibitemShut {NoStop}%
\bibitem [{\citenamefont {van~de Walle}\ and\ \citenamefont
  {Ceder}(2002{\natexlab{a}})}]{alloy_review}%
  \BibitemOpen
  \bibfield  {author} {\bibinfo {author} {\bibfnamefont {A.}~\bibnamefont
  {van~de Walle}}\ and\ \bibinfo {author} {\bibfnamefont {G.}~\bibnamefont
  {Ceder}},\ }\href {\doibase 10.1103/RevModPhys.74.11} {\bibfield  {journal}
  {\bibinfo  {journal} {Rev. Mod. Phys.}\ }\textbf {\bibinfo {volume} {74}},\
  \bibinfo {pages} {11} (\bibinfo {year} {2002}{\natexlab{a}})}\BibitemShut
  {NoStop}%
\bibitem [{\citenamefont {van~de Walle}\ and\ \citenamefont
  {Ceder}(2002{\natexlab{b}})}]{atat0}%
  \BibitemOpen
  \bibfield  {author} {\bibinfo {author} {\bibfnamefont {A.}~\bibnamefont
  {van~de Walle}}\ and\ \bibinfo {author} {\bibfnamefont {G.}~\bibnamefont
  {Ceder}},\ }\href {\doibase 10.1361/105497102770331596} {\bibfield  {journal}
  {\bibinfo  {journal} {Journal of Phase Equilibria}\ }\textbf {\bibinfo
  {volume} {23}},\ \bibinfo {pages} {348} (\bibinfo {year}
  {2002}{\natexlab{b}})}\BibitemShut {NoStop}%
\bibitem [{\citenamefont {van~de Walle}(2009)}]{atat1}%
  \BibitemOpen
  \bibfield  {author} {\bibinfo {author} {\bibfnamefont {A.}~\bibnamefont
  {van~de Walle}},\ }\href {\doibase
  https://doi.org/10.1016/j.calphad.2008.12.005} {\bibfield  {journal}
  {\bibinfo  {journal} {Calphad}\ }\textbf {\bibinfo {volume} {33}},\ \bibinfo
  {pages} {266 } (\bibinfo {year} {2009})},\ \bibinfo {note} {tools for
  Computational Thermodynamics}\BibitemShut {NoStop}%
\bibitem [{\citenamefont {Wu}\ \emph {et~al.}(2016)\citenamefont {Wu},
  \citenamefont {He}, \citenamefont {Song}, \citenamefont {Gao},\ and\
  \citenamefont {Shi}}]{cluster_review}%
  \BibitemOpen
  \bibfield  {author} {\bibinfo {author} {\bibfnamefont {Q.}~\bibnamefont
  {Wu}}, \bibinfo {author} {\bibfnamefont {B.}~\bibnamefont {He}}, \bibinfo
  {author} {\bibfnamefont {T.}~\bibnamefont {Song}}, \bibinfo {author}
  {\bibfnamefont {J.}~\bibnamefont {Gao}}, \ and\ \bibinfo {author}
  {\bibfnamefont {S.}~\bibnamefont {Shi}},\ }\href {\doibase
  https://doi.org/10.1016/j.commatsci.2016.08.034} {\bibfield  {journal}
  {\bibinfo  {journal} {Computational Materials Science}\ }\textbf {\bibinfo
  {volume} {125}},\ \bibinfo {pages} {243 } (\bibinfo {year}
  {2016})}\BibitemShut {NoStop}%
\bibitem [{\citenamefont {Ozoli\ifmmode \mbox{\c{n}}\else
  \c{n}\fi{}\ifmmode~\check{s}\else \v{s}\fi{}}\ \emph
  {et~al.}(1998{\natexlab{a}})\citenamefont {Ozoli\ifmmode \mbox{\c{n}}\else
  \c{n}\fi{}\ifmmode~\check{s}\else \v{s}\fi{}}, \citenamefont {Wolverton},\
  and\ \citenamefont {Zunger}}]{cluster_strain1}%
  \BibitemOpen
  \bibfield  {author} {\bibinfo {author} {\bibfnamefont {V.}~\bibnamefont
  {Ozoli\ifmmode \mbox{\c{n}}\else \c{n}\fi{}\ifmmode~\check{s}\else
  \v{s}\fi{}}}, \bibinfo {author} {\bibfnamefont {C.}~\bibnamefont
  {Wolverton}}, \ and\ \bibinfo {author} {\bibfnamefont {A.}~\bibnamefont
  {Zunger}},\ }\href {\doibase 10.1103/PhysRevB.57.6427} {\bibfield  {journal}
  {\bibinfo  {journal} {Phys. Rev. B}\ }\textbf {\bibinfo {volume} {57}},\
  \bibinfo {pages} {6427} (\bibinfo {year} {1998}{\natexlab{a}})}\BibitemShut
  {NoStop}%
\bibitem [{\citenamefont {Ozoli\ifmmode \mbox{\c{n}}\else
  \c{n}\fi{}\ifmmode~\check{s}\else \v{s}\fi{}}\ \emph
  {et~al.}(1998{\natexlab{b}})\citenamefont {Ozoli\ifmmode \mbox{\c{n}}\else
  \c{n}\fi{}\ifmmode~\check{s}\else \v{s}\fi{}}, \citenamefont {Wolverton},\
  and\ \citenamefont {Zunger}}]{cluster_strain2}%
  \BibitemOpen
  \bibfield  {author} {\bibinfo {author} {\bibfnamefont {V.}~\bibnamefont
  {Ozoli\ifmmode \mbox{\c{n}}\else \c{n}\fi{}\ifmmode~\check{s}\else
  \v{s}\fi{}}}, \bibinfo {author} {\bibfnamefont {C.}~\bibnamefont
  {Wolverton}}, \ and\ \bibinfo {author} {\bibfnamefont {A.}~\bibnamefont
  {Zunger}},\ }\href {\doibase 10.1103/PhysRevB.57.4816} {\bibfield  {journal}
  {\bibinfo  {journal} {Phys. Rev. B}\ }\textbf {\bibinfo {volume} {57}},\
  \bibinfo {pages} {4816} (\bibinfo {year} {1998}{\natexlab{b}})}\BibitemShut
  {NoStop}%
\bibitem [{\citenamefont {de~Gironcoli}\ \emph {et~al.}(1991)\citenamefont
  {de~Gironcoli}, \citenamefont {Giannozzi},\ and\ \citenamefont
  {Baroni}}]{sige_relax_cluster}%
  \BibitemOpen
  \bibfield  {author} {\bibinfo {author} {\bibfnamefont {S.}~\bibnamefont
  {de~Gironcoli}}, \bibinfo {author} {\bibfnamefont {P.}~\bibnamefont
  {Giannozzi}}, \ and\ \bibinfo {author} {\bibfnamefont {S.}~\bibnamefont
  {Baroni}},\ }\href {\doibase 10.1103/PhysRevLett.66.2116} {\bibfield
  {journal} {\bibinfo  {journal} {Phys. Rev. Lett.}\ }\textbf {\bibinfo
  {volume} {66}},\ \bibinfo {pages} {2116} (\bibinfo {year}
  {1991})}\BibitemShut {NoStop}%
\bibitem [{\citenamefont {Peressi}\ and\ \citenamefont
  {Baroni}(1994)}]{sige_relax_cluster2}%
  \BibitemOpen
  \bibfield  {author} {\bibinfo {author} {\bibfnamefont {M.}~\bibnamefont
  {Peressi}}\ and\ \bibinfo {author} {\bibfnamefont {S.}~\bibnamefont
  {Baroni}},\ }\href {\doibase 10.1103/PhysRevB.49.7490} {\bibfield  {journal}
  {\bibinfo  {journal} {Phys. Rev. B}\ }\textbf {\bibinfo {volume} {49}},\
  \bibinfo {pages} {7490} (\bibinfo {year} {1994})}\BibitemShut {NoStop}%
\bibitem [{\citenamefont {Garbulsky}\ and\ \citenamefont
  {Ceder}(1994)}]{cluster_vibrational}%
  \BibitemOpen
  \bibfield  {author} {\bibinfo {author} {\bibfnamefont {G.~D.}\ \bibnamefont
  {Garbulsky}}\ and\ \bibinfo {author} {\bibfnamefont {G.}~\bibnamefont
  {Ceder}},\ }\href {\doibase 10.1103/PhysRevB.49.6327} {\bibfield  {journal}
  {\bibinfo  {journal} {Phys. Rev. B}\ }\textbf {\bibinfo {volume} {49}},\
  \bibinfo {pages} {6327} (\bibinfo {year} {1994})}\BibitemShut {NoStop}%
\bibitem [{\citenamefont {Anthony}\ \emph {et~al.}(1993)\citenamefont
  {Anthony}, \citenamefont {Okamoto},\ and\ \citenamefont
  {Fultz}}]{vibrational_ni3al}%
  \BibitemOpen
  \bibfield  {author} {\bibinfo {author} {\bibfnamefont {L.}~\bibnamefont
  {Anthony}}, \bibinfo {author} {\bibfnamefont {J.~K.}\ \bibnamefont
  {Okamoto}}, \ and\ \bibinfo {author} {\bibfnamefont {B.}~\bibnamefont
  {Fultz}},\ }\href {\doibase 10.1103/PhysRevLett.70.1128} {\bibfield
  {journal} {\bibinfo  {journal} {Phys. Rev. Lett.}\ }\textbf {\bibinfo
  {volume} {70}},\ \bibinfo {pages} {1128} (\bibinfo {year}
  {1993})}\BibitemShut {NoStop}%
\bibitem [{\citenamefont {Baroni}\ \emph {et~al.}(2001)\citenamefont {Baroni},
  \citenamefont {de~Gironcoli}, \citenamefont {Dal~Corso},\ and\ \citenamefont
  {Giannozzi}}]{dft-pt}%
  \BibitemOpen
  \bibfield  {author} {\bibinfo {author} {\bibfnamefont {S.}~\bibnamefont
  {Baroni}}, \bibinfo {author} {\bibfnamefont {S.}~\bibnamefont
  {de~Gironcoli}}, \bibinfo {author} {\bibfnamefont {A.}~\bibnamefont
  {Dal~Corso}}, \ and\ \bibinfo {author} {\bibfnamefont {P.}~\bibnamefont
  {Giannozzi}},\ }\href {\doibase 10.1103/RevModPhys.73.515} {\bibfield
  {journal} {\bibinfo  {journal} {Rev. Mod. Phys.}\ }\textbf {\bibinfo {volume}
  {73}},\ \bibinfo {pages} {515} (\bibinfo {year} {2001})}\BibitemShut
  {NoStop}%
\bibitem [{\citenamefont {Esfarjani}\ and\ \citenamefont
  {Stokes}(2008)}]{keivan_anharm}%
  \BibitemOpen
  \bibfield  {author} {\bibinfo {author} {\bibfnamefont {K.}~\bibnamefont
  {Esfarjani}}\ and\ \bibinfo {author} {\bibfnamefont {H.~T.}\ \bibnamefont
  {Stokes}},\ }\href {\doibase 10.1103/PhysRevB.77.144112} {\bibfield
  {journal} {\bibinfo  {journal} {Phys. Rev. B}\ }\textbf {\bibinfo {volume}
  {77}},\ \bibinfo {pages} {144112} (\bibinfo {year} {2008})}\BibitemShut
  {NoStop}%
\bibitem [{\citenamefont {Sluiter}\ \emph {et~al.}(1999)\citenamefont
  {Sluiter}, \citenamefont {Weinert},\ and\ \citenamefont
  {Kawazoe}}]{forceconst_alloys}%
  \BibitemOpen
  \bibfield  {author} {\bibinfo {author} {\bibfnamefont {M.~H.~F.}\
  \bibnamefont {Sluiter}}, \bibinfo {author} {\bibfnamefont {M.}~\bibnamefont
  {Weinert}}, \ and\ \bibinfo {author} {\bibfnamefont {Y.}~\bibnamefont
  {Kawazoe}},\ }\href {\doibase 10.1103/PhysRevB.59.4100} {\bibfield  {journal}
  {\bibinfo  {journal} {Phys. Rev. B}\ }\textbf {\bibinfo {volume} {59}},\
  \bibinfo {pages} {4100} (\bibinfo {year} {1999})}\BibitemShut {NoStop}%
\bibitem [{\citenamefont {Zhou}\ \emph {et~al.}(2014)\citenamefont {Zhou},
  \citenamefont {Nielson}, \citenamefont {Xia},\ and\ \citenamefont
  {Ozoli\ifmmode \mbox{\c{n}}\else \c{n}\fi{}\ifmmode~\check{s}\else
  \v{s}\fi{}}}]{compressed_sensing}%
  \BibitemOpen
  \bibfield  {author} {\bibinfo {author} {\bibfnamefont {F.}~\bibnamefont
  {Zhou}}, \bibinfo {author} {\bibfnamefont {W.}~\bibnamefont {Nielson}},
  \bibinfo {author} {\bibfnamefont {Y.}~\bibnamefont {Xia}}, \ and\ \bibinfo
  {author} {\bibfnamefont {V.}~\bibnamefont {Ozoli\ifmmode \mbox{\c{n}}\else
  \c{n}\fi{}\ifmmode~\check{s}\else \v{s}\fi{}}},\ }\href {\doibase
  10.1103/PhysRevLett.113.185501} {\bibfield  {journal} {\bibinfo  {journal}
  {Phys. Rev. Lett.}\ }\textbf {\bibinfo {volume} {113}},\ \bibinfo {pages}
  {185501} (\bibinfo {year} {2014})}\BibitemShut {NoStop}%
\bibitem [{\citenamefont {Wojdeł}\ \emph {et~al.}(2013)\citenamefont
  {Wojdeł}, \citenamefont {Hermet}, \citenamefont {Ljungberg}, \citenamefont
  {Ghosez},\ and\ \citenamefont {Íñiguez}}]{lattice_dynamics}%
  \BibitemOpen
  \bibfield  {author} {\bibinfo {author} {\bibfnamefont {J.~C.}\ \bibnamefont
  {Wojdeł}}, \bibinfo {author} {\bibfnamefont {P.}~\bibnamefont {Hermet}},
  \bibinfo {author} {\bibfnamefont {M.~P.}\ \bibnamefont {Ljungberg}}, \bibinfo
  {author} {\bibfnamefont {P.}~\bibnamefont {Ghosez}}, \ and\ \bibinfo {author}
  {\bibfnamefont {J.}~\bibnamefont {Íñiguez}},\ }\href
  {http://stacks.iop.org/0953-8984/25/i=30/a=305401} {\bibfield  {journal}
  {\bibinfo  {journal} {Journal of Physics: Condensed Matter}\ }\textbf
  {\bibinfo {volume} {25}},\ \bibinfo {pages} {305401} (\bibinfo {year}
  {2013})}\BibitemShut {NoStop}%
\bibitem [{\citenamefont {Thomas}\ and\ \citenamefont
  {Ven}(2013)}]{mechanically_unstable_model}%
  \BibitemOpen
  \bibfield  {author} {\bibinfo {author} {\bibfnamefont {J.~C.}\ \bibnamefont
  {Thomas}}\ and\ \bibinfo {author} {\bibfnamefont {A.}\ \bibnamefont
  {VanderVen}},\ }\href {\doibase 10.1103/PhysRevB.88.214111} {\bibfield  {journal}
  {\bibinfo  {journal} {Phys. Rev. B}\ }\textbf {\bibinfo {volume} {88}},\
  \bibinfo {pages} {214111} (\bibinfo {year} {2013})}\BibitemShut {NoStop}%
\bibitem [{\citenamefont {Escorihuela-Sayalero}\ \emph
  {et~al.}(2017)\citenamefont {Escorihuela-Sayalero}, \citenamefont
  {Wojde\l{}},\ and\ \citenamefont {\'I\~niguez}}]{second_principles}%
  \BibitemOpen
  \bibfield  {author} {\bibinfo {author} {\bibfnamefont {C.}~\bibnamefont
  {Escorihuela-Sayalero}}, \bibinfo {author} {\bibfnamefont {J.~C.}\
  \bibnamefont {Wojde\l{}}}, \ and\ \bibinfo {author} {\bibfnamefont
  {J.}~\bibnamefont {\'I\~niguez}},\ }\href {\doibase
  10.1103/PhysRevB.95.094115} {\bibfield  {journal} {\bibinfo  {journal} {Phys.
  Rev. B}\ }\textbf {\bibinfo {volume} {95}},\ \bibinfo {pages} {094115}
  (\bibinfo {year} {2017})}\BibitemShut {NoStop}%
\bibitem [{\citenamefont {Li}\ \emph {et~al.}(2014)\citenamefont {Li},
  \citenamefont {Carrete}, \citenamefont {Katcho},\ and\ \citenamefont
  {Mingo}}]{shengbte}%
  \BibitemOpen
  \bibfield  {author} {\bibinfo {author} {\bibfnamefont {W.}~\bibnamefont
  {Li}}, \bibinfo {author} {\bibfnamefont {J.}~\bibnamefont {Carrete}},
  \bibinfo {author} {\bibfnamefont {N.~A.}\ \bibnamefont {Katcho}}, \ and\
  \bibinfo {author} {\bibfnamefont {N.}~\bibnamefont {Mingo}},\ }\href
  {\doibase https://doi.org/10.1016/j.cpc.2014.02.015} {\bibfield  {journal}
  {\bibinfo  {journal} {Computer Physics Communications}\ }\textbf {\bibinfo
  {volume} {185}},\ \bibinfo {pages} {1747 } (\bibinfo {year}
  {2014})}\BibitemShut {NoStop}%
\bibitem [{\citenamefont {Zhong}\ \emph {et~al.}(1994)\citenamefont {Zhong},
  \citenamefont {Vanderbilt},\ and\ \citenamefont {Rabe}}]{batio3_modelham}%
  \BibitemOpen
  \bibfield  {author} {\bibinfo {author} {\bibfnamefont {W.}~\bibnamefont
  {Zhong}}, \bibinfo {author} {\bibfnamefont {D.}~\bibnamefont {Vanderbilt}}, \
  and\ \bibinfo {author} {\bibfnamefont {K.~M.}\ \bibnamefont {Rabe}},\ }\href
  {\doibase 10.1103/PhysRevLett.73.1861} {\bibfield  {journal} {\bibinfo
  {journal} {Phys. Rev. Lett.}\ }\textbf {\bibinfo {volume} {73}},\ \bibinfo
  {pages} {1861} (\bibinfo {year} {1994})}\BibitemShut {NoStop}%
\bibitem [{\citenamefont {Zhong}\ \emph {et~al.}(1995)\citenamefont {Zhong},
  \citenamefont {Vanderbilt},\ and\ \citenamefont {Rabe}}]{batio3_modelham2}%
  \BibitemOpen
  \bibfield  {author} {\bibinfo {author} {\bibfnamefont {W.}~\bibnamefont
  {Zhong}}, \bibinfo {author} {\bibfnamefont {D.}~\bibnamefont {Vanderbilt}}, \
  and\ \bibinfo {author} {\bibfnamefont {K.~M.}\ \bibnamefont {Rabe}},\ }\href
  {\doibase 10.1103/PhysRevB.52.6301} {\bibfield  {journal} {\bibinfo
  {journal} {Phys. Rev. B}\ }\textbf {\bibinfo {volume} {52}},\ \bibinfo
  {pages} {6301} (\bibinfo {year} {1995})}\BibitemShut {NoStop}%
\bibitem [{\citenamefont {Zhong}\ and\ \citenamefont
  {Vanderbilt}(1996)}]{modelham_quantum}%
  \BibitemOpen
  \bibfield  {author} {\bibinfo {author} {\bibfnamefont {W.}~\bibnamefont
  {Zhong}}\ and\ \bibinfo {author} {\bibfnamefont {D.}~\bibnamefont
  {Vanderbilt}},\ }\href {\doibase 10.1103/PhysRevB.53.5047} {\bibfield
  {journal} {\bibinfo  {journal} {Phys. Rev. B}\ }\textbf {\bibinfo {volume}
  {53}},\ \bibinfo {pages} {5047} (\bibinfo {year} {1996})}\BibitemShut
  {NoStop}%
\bibitem [{\citenamefont {Bellaiche}\ \emph {et~al.}(2000)\citenamefont
  {Bellaiche}, \citenamefont {Garc\'{\i}a},\ and\ \citenamefont
  {Vanderbilt}}]{pzt_modelham}%
  \BibitemOpen
  \bibfield  {author} {\bibinfo {author} {\bibfnamefont {L.}~\bibnamefont
  {Bellaiche}}, \bibinfo {author} {\bibfnamefont {A.}~\bibnamefont
  {Garc\'{\i}a}}, \ and\ \bibinfo {author} {\bibfnamefont {D.}~\bibnamefont
  {Vanderbilt}},\ }\href {\doibase 10.1103/PhysRevLett.84.5427} {\bibfield
  {journal} {\bibinfo  {journal} {Phys. Rev. Lett.}\ }\textbf {\bibinfo
  {volume} {84}},\ \bibinfo {pages} {5427} (\bibinfo {year}
  {2000})}\BibitemShut {NoStop}%
\bibitem [{\citenamefont {Walizer}\ \emph {et~al.}(2006)\citenamefont
  {Walizer}, \citenamefont {Lisenkov},\ and\ \citenamefont
  {Bellaiche}}]{bst_modelham}%
  \BibitemOpen
  \bibfield  {author} {\bibinfo {author} {\bibfnamefont {L.}~\bibnamefont
  {Walizer}}, \bibinfo {author} {\bibfnamefont {S.}~\bibnamefont {Lisenkov}}, \
  and\ \bibinfo {author} {\bibfnamefont {L.}~\bibnamefont {Bellaiche}},\ }\href
  {\doibase 10.1103/PhysRevB.73.144105} {\bibfield  {journal} {\bibinfo
  {journal} {Phys. Rev. B}\ }\textbf {\bibinfo {volume} {73}},\ \bibinfo
  {pages} {144105} (\bibinfo {year} {2006})}\BibitemShut {NoStop}%
\bibitem [{\citenamefont {Leibfried}\ and\ \citenamefont
  {Ludwig}(1961)}]{anharmonic}%
  \BibitemOpen
  \bibfield  {author} {\bibinfo {author} {\bibfnamefont {G.}~\bibnamefont
  {Leibfried}}\ and\ \bibinfo {author} {\bibfnamefont {W.}~\bibnamefont
  {Ludwig}}\ }(\bibinfo  {publisher} {Academic Press},\ \bibinfo {year}
  {1961})\ pp.\ \bibinfo {pages} {275 -- 444}\BibitemShut {NoStop}%
\bibitem [{kun(1950)}]{kun_huang}%
  \BibitemOpen
  \href {\doibase 10.1098/rspa.1950.0133} {\bibfield  {journal} {\bibinfo
  {journal} {Proceedings of the Royal Society of London A: Mathematical,
  Physical and Engineering Sciences}\ }\textbf {\bibinfo {volume} {203}},\
  \bibinfo {pages} {178} (\bibinfo {year} {1950})}\BibitemShut {NoStop}%
\bibitem [{\citenamefont {Guyon}\ \emph {et~al.}(2002)\citenamefont {Guyon},
  \citenamefont {Weston}, \citenamefont {Barnhill},\ and\ \citenamefont
  {Vapnik}}]{recursive_fe}%
  \BibitemOpen
  \bibfield  {author} {\bibinfo {author} {\bibfnamefont {I.}~\bibnamefont
  {Guyon}}, \bibinfo {author} {\bibfnamefont {J.}~\bibnamefont {Weston}},
  \bibinfo {author} {\bibfnamefont {S.}~\bibnamefont {Barnhill}}, \ and\
  \bibinfo {author} {\bibfnamefont {V.}~\bibnamefont {Vapnik}},\ }\href
  {\doibase 10.1023/A:1012487302797} {\bibfield  {journal} {\bibinfo  {journal}
  {Machine Learning}\ }\textbf {\bibinfo {volume} {46}},\ \bibinfo {pages}
  {389} (\bibinfo {year} {2002})}\BibitemShut {NoStop}%
\bibitem [{\citenamefont {Gonze}\ and\ \citenamefont {Lee}(1997)}]{gonze}%
  \BibitemOpen
  \bibfield  {author} {\bibinfo {author} {\bibfnamefont {X.}~\bibnamefont
  {Gonze}}\ and\ \bibinfo {author} {\bibfnamefont {C.}~\bibnamefont {Lee}},\
  }\href {\doibase 10.1103/PhysRevB.55.10355} {\bibfield  {journal} {\bibinfo
  {journal} {Phys. Rev. B}\ }\textbf {\bibinfo {volume} {55}},\ \bibinfo
  {pages} {10355} (\bibinfo {year} {1997})}\BibitemShut {NoStop}%
\bibitem [{\citenamefont {Hohenberg}\ and\ \citenamefont {Kohn}(1964)}]{hk}%
  \BibitemOpen
  \bibfield  {author} {\bibinfo {author} {\bibfnamefont {P.}~\bibnamefont
  {Hohenberg}}\ and\ \bibinfo {author} {\bibfnamefont {W.}~\bibnamefont
  {Kohn}},\ }\href@noop {} {\bibfield  {journal} {\bibinfo  {journal} {Phys.\
  Rev.}\ }\textbf {\bibinfo {volume} {136}},\ \bibinfo {pages} {B864} (\bibinfo
  {year} {1964})}\BibitemShut {NoStop}%
\bibitem [{\citenamefont {Kohn}\ and\ \citenamefont {Sham}(1965)}]{ks}%
  \BibitemOpen
  \bibfield  {author} {\bibinfo {author} {\bibfnamefont {W.}~\bibnamefont
  {Kohn}}\ and\ \bibinfo {author} {\bibfnamefont {L.}~\bibnamefont {Sham}},\
  }\href@noop {} {\bibfield  {journal} {\bibinfo  {journal} {Phys.\ Rev.}\
  }\textbf {\bibinfo {volume} {140}},\ \bibinfo {pages} {A1133} (\bibinfo
  {year} {1965})}\BibitemShut {NoStop}%
\bibitem [{\citenamefont {Giannozzi}\ and\ \citenamefont {et~al.}(2009)}]{QE}%
  \BibitemOpen
  \bibfield  {author} {\bibinfo {author} {\bibfnamefont {P.}~\bibnamefont
  {Giannozzi}}\ and\ \bibinfo {author} {\bibnamefont {et~al.}},\ }\href@noop {}
  {\bibfield  {journal} {\bibinfo  {journal} {J. Phys.:Condens. Matter}\
  }\textbf {\bibinfo {volume} {21}},\ \bibinfo {pages} {395502} (\bibinfo
  {year} {2009})}\BibitemShut {NoStop}%
\bibitem [{\citenamefont {Vanderbilt}(1990)}]{ultrasoft}%
  \BibitemOpen
  \bibfield  {author} {\bibinfo {author} {\bibfnamefont {D.}~\bibnamefont
  {Vanderbilt}},\ }\href@noop {} {\bibfield  {journal} {\bibinfo  {journal}
  {Phys. Rev. B}\ }\textbf {\bibinfo {volume} {41}},\ \bibinfo {pages} {7892}
  (\bibinfo {year} {1990})}\BibitemShut {NoStop}%
\bibitem [{\citenamefont {Garrity}\ \emph {et~al.}(2014)\citenamefont
  {Garrity}, \citenamefont {Bennett}, \citenamefont {Rabe},\ and\ \citenamefont
  {Vanderbilt}}]{gbrv}%
  \BibitemOpen
  \bibfield  {author} {\bibinfo {author} {\bibfnamefont {K.~F.}\ \bibnamefont
  {Garrity}}, \bibinfo {author} {\bibfnamefont {J.~W.}\ \bibnamefont
  {Bennett}}, \bibinfo {author} {\bibfnamefont {K.~M.}\ \bibnamefont {Rabe}}, \
  and\ \bibinfo {author} {\bibfnamefont {D.}~\bibnamefont {Vanderbilt}},\
  }\href@noop {} {\bibfield  {journal} {\bibinfo  {journal} {Comput. Mater.
  Sci}\ }\textbf {\bibinfo {volume} {81}},\ \bibinfo {pages} {446} (\bibinfo
  {year} {2014})}\BibitemShut {NoStop}%
\bibitem [{\citenamefont {Perdew}\ \emph {et~al.}(2008)\citenamefont {Perdew},
  \citenamefont {Ruzsinszky}, \citenamefont {Csonka}, \citenamefont {Vydrov},
  \citenamefont {Scuseria}, \citenamefont {Constantin}, \citenamefont {Zhou},\
  and\ \citenamefont {Burke}}]{pbesol}%
  \BibitemOpen
  \bibfield  {author} {\bibinfo {author} {\bibfnamefont {J.~P.}\ \bibnamefont
  {Perdew}}, \bibinfo {author} {\bibfnamefont {A.}~\bibnamefont {Ruzsinszky}},
  \bibinfo {author} {\bibfnamefont {G.~I.}\ \bibnamefont {Csonka}}, \bibinfo
  {author} {\bibfnamefont {O.~A.}\ \bibnamefont {Vydrov}}, \bibinfo {author}
  {\bibfnamefont {G.~E.}\ \bibnamefont {Scuseria}}, \bibinfo {author}
  {\bibfnamefont {L.~A.}\ \bibnamefont {Constantin}}, \bibinfo {author}
  {\bibfnamefont {X.}~\bibnamefont {Zhou}}, \ and\ \bibinfo {author}
  {\bibfnamefont {K.}~\bibnamefont {Burke}},\ }\href {\doibase
  10.1103/PhysRevLett.100.136406} {\bibfield  {journal} {\bibinfo  {journal}
  {Phys. Rev. Lett.}\ }\textbf {\bibinfo {volume} {100}},\ \bibinfo {pages}
  {136406} (\bibinfo {year} {2008})}\BibitemShut {NoStop}%
\bibitem [{\citenamefont {Lloyd-Williams}\ and\ \citenamefont
  {Monserrat}(2015)}]{nondiag}%
  \BibitemOpen
  \bibfield  {author} {\bibinfo {author} {\bibfnamefont {J.~H.}\ \bibnamefont
  {Lloyd-Williams}}\ and\ \bibinfo {author} {\bibfnamefont {B.}~\bibnamefont
  {Monserrat}},\ }\href {\doibase 10.1103/PhysRevB.92.184301} {\bibfield
  {journal} {\bibinfo  {journal} {Phys. Rev. B}\ }\textbf {\bibinfo {volume}
  {92}},\ \bibinfo {pages} {184301} (\bibinfo {year} {2015})}\BibitemShut
  {NoStop}%
\bibitem [{\citenamefont {Hastings}(1970)}]{montecarlo}%
  \BibitemOpen
  \bibfield  {author} {\bibinfo {author} {\bibfnamefont {W.~K.}\ \bibnamefont
  {Hastings}},\ }\href {\doibase 10.1093/biomet/57.1.97} {\bibfield  {journal}
  {\bibinfo  {journal} {Biometrika}\ }\textbf {\bibinfo {volume} {57}},\
  \bibinfo {pages} {97} (\bibinfo {year} {1970})}\BibitemShut {NoStop}%
\bibitem [{\citenamefont {Qteish}\ and\ \citenamefont
  {Resta}(1988)}]{sige_old}%
  \BibitemOpen
  \bibfield  {author} {\bibinfo {author} {\bibfnamefont {A.}~\bibnamefont
  {Qteish}}\ and\ \bibinfo {author} {\bibfnamefont {R.}~\bibnamefont {Resta}},\
  }\href {\doibase 10.1103/PhysRevB.37.6983} {\bibfield  {journal} {\bibinfo
  {journal} {Phys. Rev. B}\ }\textbf {\bibinfo {volume} {37}},\ \bibinfo
  {pages} {6983} (\bibinfo {year} {1988})}\BibitemShut {NoStop}%
\bibitem [{\citenamefont {Paddison}\ \emph {et~al.}(2018)\citenamefont
  {Paddison}, \citenamefont {Gutmann}, \citenamefont {Stewart}, \citenamefont
  {Tucker}, \citenamefont {Dove}, \citenamefont {Keen},\ and\ \citenamefont
  {Goodwin}}]{mno}%
  \BibitemOpen
  \bibfield  {author} {\bibinfo {author} {\bibfnamefont {J.~A.~M.}\
  \bibnamefont {Paddison}}, \bibinfo {author} {\bibfnamefont {M.~J.}\
  \bibnamefont {Gutmann}}, \bibinfo {author} {\bibfnamefont {J.~R.}\
  \bibnamefont {Stewart}}, \bibinfo {author} {\bibfnamefont {M.~G.}\
  \bibnamefont {Tucker}}, \bibinfo {author} {\bibfnamefont {M.~T.}\
  \bibnamefont {Dove}}, \bibinfo {author} {\bibfnamefont {D.~A.}\ \bibnamefont
  {Keen}}, \ and\ \bibinfo {author} {\bibfnamefont {A.~L.}\ \bibnamefont
  {Goodwin}},\ }\href {\doibase 10.1103/PhysRevB.97.014429} {\bibfield
  {journal} {\bibinfo  {journal} {Phys. Rev. B}\ }\textbf {\bibinfo {volume}
  {97}},\ \bibinfo {pages} {014429} (\bibinfo {year} {2018})}\BibitemShut
  {NoStop}%
\bibitem [{\citenamefont {Turner}\ \emph {et~al.}(1997)\citenamefont {Turner},
  \citenamefont {Zhu}, \citenamefont {Chan},\ and\ \citenamefont {Ho}}]{al}%
  \BibitemOpen
  \bibfield  {author} {\bibinfo {author} {\bibfnamefont {D.~E.}\ \bibnamefont
  {Turner}}, \bibinfo {author} {\bibfnamefont {Z.~Z.}\ \bibnamefont {Zhu}},
  \bibinfo {author} {\bibfnamefont {C.~T.}\ \bibnamefont {Chan}}, \ and\
  \bibinfo {author} {\bibfnamefont {K.~M.}\ \bibnamefont {Ho}},\ }\href
  {\doibase 10.1103/PhysRevB.55.13842} {\bibfield  {journal} {\bibinfo
  {journal} {Phys. Rev. B}\ }\textbf {\bibinfo {volume} {55}},\ \bibinfo
  {pages} {13842} (\bibinfo {year} {1997})}\BibitemShut {NoStop}%
\bibitem [{\citenamefont {Glazer}(1972)}]{glazer}%
  \BibitemOpen
  \bibfield  {author} {\bibinfo {author} {\bibfnamefont {A.~M.}\ \bibnamefont
  {Glazer}},\ }\href {\doibase 10.1107/S0567740872007976} {\bibfield  {journal}
  {\bibinfo  {journal} {Acta Crystallographica Section B}\ }\textbf {\bibinfo
  {volume} {28}},\ \bibinfo {pages} {3384} (\bibinfo {year}
  {1972})}\BibitemShut {NoStop}%
\bibitem [{\citenamefont {Sai}\ and\ \citenamefont
  {Vanderbilt}(2000)}]{sto_rotations}%
  \BibitemOpen
  \bibfield  {author} {\bibinfo {author} {\bibfnamefont {N.}~\bibnamefont
  {Sai}}\ and\ \bibinfo {author} {\bibfnamefont {D.}~\bibnamefont
  {Vanderbilt}},\ }\href {\doibase 10.1103/PhysRevB.62.13942} {\bibfield
  {journal} {\bibinfo  {journal} {Phys. Rev. B}\ }\textbf {\bibinfo {volume}
  {62}},\ \bibinfo {pages} {13942} (\bibinfo {year} {2000})}\BibitemShut
  {NoStop}%
\end{thebibliography}
%

\end{document}